\documentclass[prd, twocolumn, superscriptaddress,floatfix, nofootinbib, preprintnumbers]{revtex4-2}
\usepackage[utf8]{inputenc}
\usepackage{hyperref}
\hypersetup{colorlinks=true,citecolor=blue}
\usepackage{graphicx}
\usepackage[caption=false]{subfig}
\usepackage{amsmath,amssymb,amstext}
\usepackage{bm}
\usepackage{booktabs}
\usepackage{multirow}
\usepackage{enumitem}
\usepackage{physics}
\usepackage{tikz}
\usepackage{float}

\usepackage{diagbox}

\renewcommand{\t}[1]{{\tt #1}}

\begin{document}

\title{SURFing to the Fundamental Limit of Jet Tagging}

\author{Ian Pang}
\email{ian.pang@physics.rutgers.edu}
\affiliation{NHETC, Dept.\ of Physics and Astronomy, Rutgers University, Piscataway, NJ 08854, USA}
\author{Darius A. Faroughy}
\email{darius.faroughy@rutgers.edu}
\affiliation{NHETC, Dept.\ of Physics and Astronomy, Rutgers University, Piscataway, NJ 08854, USA}
\author{David Shih}
\email{shih@physics.rutgers.edu}
\affiliation{NHETC, Dept.\ of Physics and Astronomy, Rutgers University, Piscataway, NJ 08854, USA}
\author{Ranit Das}
\email{das@thphys.uni-heidelberg.de
}
\affiliation{NHETC, Dept.\ of Physics and Astronomy, Rutgers University, Piscataway, NJ 08854, USA}
\affiliation{Institute for Theoretical Physics, Universit\"{a}t Heidelberg, Germany}
\author{Gregor Kasieczka}
\email{gregor.kasieczka@uni-hamburg.de}
\affiliation{Institut f\"{u}r Experimentalphysik, Universit\"{a}t Hamburg, 22761 Hamburg, Germany}

\begin{abstract}

Beyond the practical goal of improving search and measurement sensitivity through better jet tagging algorithms, there is a deeper question: what are their upper performance limits? Generative surrogate models with learned likelihood functions offer a new approach to this problem, provided the surrogate correctly captures the underlying data distribution. 
In this work, we introduce the SUrrogate ReFerence (SURF) method, a new approach to validating generative models. This framework enables exact Neyman–Pearson tests by training the target model on samples from another tractable surrogate, which is itself trained on real data. We argue that the 
EPiC-FM generative model is a valid surrogate reference for \textsc{JetClass} jets and apply SURF to show that modern jet taggers may already be operating close to the true statistical limit. By contrast, we find that autoregressive GPT models unphysically exaggerate top vs. QCD separation power encoded in the surrogate reference, implying that they are giving a misleading picture of the fundamental limit.

\end{abstract}

\maketitle

\section{Introduction}

Jet tagging is a central task in collider physics. Over the past decade, machine learning has driven major advances in jet tagging, with increasingly sophisticated architectures achieving very high classification performance on simulated datasets~\cite{Qu:2019gqs,Kasieczka:2019dbj,CMS:2020poo,Gong:2022lye,Bogatskiy:2023nnw,Qu:2022mxj,Brehmer:2024yqw,Mikuni:2024qsr,Mikuni:2025tar,ATLAS:2024rua,Favaro:2025pgz,Bhimji:2025isp}. This success naturally raises a key question: \emph{have current jet taggers already reached the fundamental limit of jet tagging, or does a gap remain between practical performance and the true statistical optimum?}

The Neyman-Pearson (NP) limit, defined by the likelihood ratio, is the best possible discriminant between two different underlying physics processes -- such as top and QCD jets -- that any classifier could achieve if it had access to the exact data likelihoods~\cite{Neyman:1933wgr}. In practice, however, this limit cannot be evaluated directly because the true likelihood of the data-generating process is unknown. It therefore remains unclear how close existing classifiers are to this ultimate bound.

Recently, Ref.~\cite{Geuskens:2024tfo} proposed using autoregressive GPT-style generative models to probe this limit for top vs.\ QCD jets from the \textsc{JetClass} dataset~\cite{JetClass}. These models operate on discretized, tokenized representations of jet constituents and yield explicit log-likelihoods, enabling the computation of likelihood ratios between jet classes. When applied to top--QCD discrimination, GPT-based likelihoods were found to produce receiver-operator-characteristic (ROC) curves that appear to substantially exceed those of all existing classifiers trained on GPT samples. This was interpreted as evidence that current jet taggers fall well short of the true NP limit.

In this work, we reevaluate this conclusion in several ways. First, we  upgrade the GPT model of Refs.~\cite{Finke:2023veq,Geuskens:2024tfo} with a number of enhancements, including voxel tokenization, positional encoding and weight tying. 
Despite this, we find that the improved GPT model trained on tops and QCD from \textsc{JetClass} still inflates their separation relative to state-of-the-art classifiers by more than an order of magnitude in rejection factor over a wide range of signal efficiencies. In fact, this is an even larger gap than the original one of Refs.~\cite{Finke:2023veq,Geuskens:2024tfo}. Along the way, we show that GPT-generated jets are perfectly separable from continuous jets, while being close to the discretized jets that the GPT models were trained on. However, we show this cannot explain the inflated separation of the GPT jets -- the discretized top and QCD jets are if anything {\it less} separable than their continuous counterparts.

Secondly, we assess the fundamental limit of jet tagging using an alternative generative modeling framework that also gives tractable likelihoods: conditional flow matching (CFM) \cite{albergo2022building, lipman2022flow, tong2023improving} with permutation equivariance, in the form of the EPiC-FM architecture~\cite{Buhmann:2023zgc,Birk:2023efj}. Unlike the GPT framework, EPiC-FM operates directly at the level of the continuous jet constituents and does not require any discretization or tokenization.
Using the log-likelihood ratios derived from EPiC-FM trained on \textsc{JetClass} QCD and top jets, we find no evidence for an inflated fundamental limit: the ROC curve obtained from the exact NP test is fairly close (within a factor of two) to the state-of-the-art classifier performance on EPiC-FM samples.

These results raise the question -- which jet generative model is giving the more accurate picture of the fundamental limit of jet tagging? Is there an enormous gap between the fundamental limit and our best classifiers, as the GPT models would suggest? Or is there essentially no gap between the two, as the EPiC-FM models would suggest?

To clarify the origin of this discrepancy and to provide a general framework for evaluating generative models, we introduce the SUrrogate ReFerence (SURF) method. Here the key idea is to use one generative model with tractable likelihoods (EPiC-FM) to evaluate the other generative model with tractable likelihoods (GPT). The first one serves as the {\it surrogate reference model} for the second. By various ways we validate EPiC-FM as a surrogate reference model, checking that it does not introduce spurious artifacts into the \textsc{JetClass} data that are hard to detect.\footnote{As we explain in detail in Sec.~\ref{sec:gpt_model}, the same validation tests fail for the GPT model, indicating that it does not satisfy the requirements of a reliable surrogate reference. 
For this reason, we do not recommend running the SURF method in the opposite order.} We replace \textsc{JetClass} samples with 
samples drawn from EPiC-FM, train the GPT model on these samples, and repeat all the studies and tests we want to do with the GPT model. But now with tractable likelihoods for both the surrogate reference and the target generative model, ground truth NP-optimal ROC curves can be calculated and compared against one another, giving us more information than when we had trained the GPT model on \textsc{JetClass} directly. 

Within this framework, we find that the NP ROC curve derived from the GPT likelihoods far exceeds that of the EPiC-FM surrogate itself. This demonstrates unambiguously that the GPT model artificially inflates the apparent separation between top and QCD jets beyond the physical limit represented by the surrogate reference. Evidently, the GPT model is introducing unphysical artifacts into the EPiC-FM surrogate samples that it is trained on, leading them to be more separable than they should be. We investigate the reasons for this and find evidence that GPT models exhibit signs of overfitting to their training data. We speculate how this form of mismodeling might produce likelihood ratios that are inherently difficult for classifiers to reproduce, leading to artificially inflated ROC curves.

In the end, we find no evidence of an inflated ``fundamental limit" of jet tagging.
Instead, the overall picture, especially the concurrence of top vs.\ QCD ROC curves from OmniLearn trained on \textsc{JetClass} and on EPiC-FM together with the exact likelihoods of EPiC-FM, suggests that the fundamental limit is not far away from state-of-the-art classifiers.

The outline of our paper is as follows. In Sections~\ref{sec:gpt_model} and~\ref{sec:epicfm}, we present the model descriptions, baseline evaluations, and top vs.\ QCD separation results obtained using the GPT model and the EPiC-FM model, respectively. Section~\ref{sec:surrogate_method} introduces the SURF method and its corresponding results. Possible reasons for the inflated GPT top vs.\ QCD curve are examined in Section~\ref{sec:roc_inflation}. 
We conclude with an outlook in Section~\ref{sec:conclusion}, while additional technical background and log-likelihood details are provided in the appendices.

\section{GPT: Jets as tokenized sequences}
\label{sec:gpt_model}

\subsection{Description of the model} 
\label{subsec:gpt_desc}
One popular generative modeling framework for jets is to view them as $p_T$-ordered sequences of tokenized particles. We follow the approach of Ref.~\cite{Geuskens:2024tfo}, where the kinematic features of each constituent are discretized into bins and mapped to tokens. The resulting sequences can then be modeled autoregressively with GPT-style autoregressive language models, which learn the joint distribution over jet constituents by factorizing it into a product of conditional probabilities for each tokenized particle given the preceding ones. We note that VQ-VAE–based tokenization is also possible~\cite{Golling:2024abg,Birk:2024knn}; however, obtaining a well-defined likelihood in physical space is nontrivial because the learned codebook partitions feature space into irregular regions without a known volume element. Establishing a principled mapping from sequence log-probabilities to physical-space log-densities in that setting is left to future work, so we adopt a fixed-bin vocabulary here.

Concretely, within each jet $\mathbf{x}=\{\mathbf{x}_i\}_{i=1}^{N}$ we order the constituents $\mathbf{x}_i=( \log   p_T,\Delta\eta,\Delta\phi)_i$ by descending $p_T$ and discretize the kinematic features into $N_{p_T}=40$, $N_{\Delta\eta}=30$, and $N_{\Delta\phi}=30$ evenly spaced bins. Here $p_{T,i}$ is the transverse momenta of particle $i$, $\Delta\eta_i\equiv \eta_i-\eta_J$ is the pseudorapidity offset from the jet axis, and $\Delta\phi_i\equiv\phi_i-\phi_J$ is the azimuthal-angle offset. Each three-dimensional voxel in feature space is mapped to a unique token, giving rise to a vocabulary of size $V=N_{p_T}\cdot N_{\Delta\eta}\cdot N_{\Delta\phi}=36{,}000$. We restrict each jet to the 40 highest-$p_T$ particles before discretization.\footnote{We have also performed a minimal study using up to 128 particles per jet and observe consistent behavior, with no qualitative change in our conclusions. These results are included in appendix~\ref{app:128particles}} This choice is motivated by Ref.~\cite{Geuskens:2024tfo}, which observed a sizable gap between the NP ROC curve and that of a trained classifier even under this 40-particle restriction. Increasing the number of particles only slightly enlarged the gap without altering the conclusion, and thus does not justify the additional computational cost. The pseudo-rapidity and azimuthal bins are restricted to the range $(-0.8,\,0.8)$ with overflows. Each jet $\mathbf{x}$ is thus represented by a sequence ${\t{\mathbf t}_\mathbf{x}}=(\t t_1,\dots, \t t_N)$, where each token $\t t_i$ represents an individual particle. Sequences are padded to a maximum length of $N_{\max}=40$ using a dedicated \texttt{[PAD]} token. Additional special tokens, \texttt{[START]} and \texttt{[END]}, are appended to the vocabulary to indicate the beginning and end of each jet sequence. 

We implement our GPT model with the HuggingFace Transformers library~\cite{wolf2020transformers}. Model and training details can be found in appendix~\ref{app:models-gpt}. When compared to Ref.~\cite{Geuskens:2024tfo}, our setup differs in three key ways.
(i) \emph{Voxel tokenization:} we use a single, joint vocabulary over $(\log p_T,\Delta\eta,\Delta\phi)$ voxels and embed each token directly into $\mathbb{R}^{d_{\rm embd}}$, whereas Ref.~\cite{Geuskens:2024tfo} tokenizes each binned feature separately and sums three feature-specific embeddings to form the per-particle representation.
(ii) \emph{Positional information:} we add learned positional embeddings (reflecting the $p_T$ ordering) to the token embeddings before the self-attention layers; Ref.~\cite{Geuskens:2024tfo} omits positional encodings.
(iii) \emph{Weight tying:} we tie the last linear layer (language head) to the input token embedding matrix, using the same weights for input embedding and output logits. Weight tying reduces the number of trainable parameters and has been shown to improve generalization in language models~\cite{press2017using,inan2017tying}. Empirically, we find these choices improve the generative performance of our GPT model.

Drawing a sample from a trained GPT produces tokens $\t{\mathbf  t}_\mathbf{x}=(\t t_0={\tt [START]},\t t_1,\dots,\t t_{N}, \t t_{N+1}={\tt [END]})$ which correspond to binned, $p_T$-ordered jet constituents. The log-likelihood of a tokenized jet  is computed by summing the next-token probabilities 
\begin{equation}
\label{eq:gpt_ll}
  \log p_\theta({\t{\mathbf t}_\mathbf{x}})
  = \sum_{n=1}^{N+1} \log p_\theta\!\left(\t t_n\,\big|\,\t t_0,\dots,\t t_{n-1}\right),
\end{equation}
along the sequence and stopping at the {\tt [END]} symbol.

The tokenized jets can be mapped back to the physical space of continuous jet constituents by identifying each constituent token with its corresponding voxel in $(\log p_T,\Delta\eta,\Delta\phi)$ space and then smearing it within its voxel according to a uniform density. The corresponding log-likelihood in the physical space is then given by 
\begin{equation}\label{eq:logpcont}
    \log p^{cont}_\theta(\mathbf{x}) =\log p_\theta(\t{\mathbf t}_\mathbf{x})- N\log dV+ \sum_k \log N_k! - \log N!
\end{equation}
Here the second term is the correction to the log-likelihood from the uniform smearing ($dV$ is the volume of the voxel), and the third and fourth terms are necessary to remove the spurious effects of $p_T$-ordering in the sequence ($N_k$ is the number of constituents with identical discretized $p_T$).

Clearly, with the GPT models there are two choices for what we consider as the ``original data" -- we can use either the original continuous \textsc{JetClass} samples or their bin-smeared counterparts as the baseline. Ostensibly only the former case is of interest -- we are ultimately interested in generative models for continuous jets. However, the latter is also a useful comparison point, and we will consider both in the following.

\subsection{Baseline evaluation of the model}
\label{sec:validate_baseline}

We first perform a baseline evaluation of GPT models trained directly on \textsc{JetClass} data. For brevity, we evaluate performance solely using the AUC of a classifier trained to distinguish generated samples from real data~\cite{Krause:2021ilc,Krause:2024avx}, although other evaluation strategies are possible~\cite{Lim:2022nft,Kansal:2022spb,Das:2023ktd,Cappelli:2025myc}.

We train an OmniLearn classifier\footnote{For all our results, we use OmniLearn from Refs.~\cite{Mikuni:2024qsr,Mikuni:2025tar} to represent current state-of-the-art classifiers. We pretrain it on the entire \textsc{JetClass} training dataset and fine-tune it on the jets used in our study. In particular, we use 1M jets from each class, with an 80/10/10 split into training, validation, and test sets.} 
to separate \textsc{JetClass} data from GPT-generated jets. 
We find that OmniLearn separates both datasets almost perfectly (AUC $=0.9949$ for QCD  and $0.9734$ for tops; Table~\ref{tab:aucs_baseline}, first row). The near-perfect separability indicates that GPT samples lie far from the continuous data manifold.

\begin{table}
\centering
\renewcommand{\arraystretch}{2.5}
\begin{tabular}{|l|c|c|}
\hline
\textbf{Comparison} & \textbf{QCD} & \textbf{Top} \\ \hline
\textsc{JetClass} vs GPT
  & \begin{tikzpicture}[baseline=(current bounding box.center)]
      \node[minimum width=2cm, minimum height=1.5cm] (box) {};
      \draw (box.south west) -- (box.north east);
      \node at ([shift={(-0.7cm,0.2cm)}]box.center) {0.9949};
      \node at ([shift={(0.4cm,-0.2cm)}]box.center) {?};
    \end{tikzpicture}
  & \begin{tikzpicture}[baseline=(current bounding box.center)]
      \node[minimum width=2cm, minimum height=1.5cm] (box) {};
      \draw (box.south west) -- (box.north east);
      \node at ([shift={(-0.7cm,0.2cm)}]box.center) {0.9734};
      \node at ([shift={(0.4cm,-0.2cm)}]box.center) {?};
    \end{tikzpicture} \\ \hline
\textsc{JetClass}-bs vs GPT
  & \begin{tikzpicture}[baseline=(current bounding box.center)]
      \node[minimum width=2cm, minimum height=1.5cm] (box) {};
      \draw (box.south west) -- (box.north east);
      \node at ([shift={(-0.7cm,0.2cm)}]box.center) {0.5564};
      \node at ([shift={(0.4cm,-0.2cm)}]box.center) {?};
    \end{tikzpicture}
  & \begin{tikzpicture}[baseline=(current bounding box.center)]
      \node[minimum width=2cm, minimum height=1.5cm] (box) {};
      \draw (box.south west) -- (box.north east);
      \node at ([shift={(-0.7cm,0.2cm)}]box.center) {0.5965};
      \node at ([shift={(0.4cm,-0.2cm)}]box.center) {?};
    \end{tikzpicture} \\ \hline
\textsc{JetClass} vs \textsc{JetClass}-bs
  & \begin{tikzpicture}[baseline=(current bounding box.center)]
      \node[minimum width=2cm, minimum height=1.5cm] (box) {};
      \draw (box.south west) -- (box.north east);
      \node at ([shift={(-0.7cm,0.2cm)}]box.center) {0.9940};
      \node at ([shift={(0.4cm,-0.2cm)}]box.center) {?};
    \end{tikzpicture}
  & \begin{tikzpicture}[baseline=(current bounding box.center)]
      \node[minimum width=2cm, minimum height=1.5cm] (box) {};
      \draw (box.south west) -- (box.north east);
      \node at ([shift={(-0.7cm,0.2cm)}]box.center) {0.9714};
      \node at ([shift={(0.4cm,-0.2cm)}]box.center) {?};
    \end{tikzpicture} \\ \hline
\end{tabular}

\caption{
AUC scores for distinguishing between jets from \textsc{JetClass}, bin-smeared \textsc{JetClass}, and GPT trained on \textsc{JetClass}, 
using the OmniLearn classifier. Below the diagonal line the space is reserved for the true NP-optimal classifier scores, which cannot be computed because \textsc{JetClass} likelihoods are unavailable.
}
\label{tab:aucs_baseline}
\end{table}

Is this due to an issue with the GPT models, or is it due to the binning of the \textsc{JetClass} data itself? We answer this question with 
two more classifier tests: GPT samples vs.\ bin-smeared \textsc{JetClass}, and continuous \textsc{JetClass} vs.\ bin-smeared \textsc{JetClass}. The first is shown in 
Table~\ref{tab:aucs_baseline}, second row, indicating OmniLearn AUCs of 0.5564 and 0.5965 for QCD and tops respectively. 
This reproduces the good performance achieved by the original Aachen GPT-style jet generator~\cite{Finke:2023veq} and in fact improves upon it.\footnote{Retraining OmniLearn on bin-smeared Aachen GPT–generated jets from Ref.~\cite{reyes_gonzalez_2024_14023638} vs.\ bin-smeared \textsc{JetClass} jets yields AUC values of 0.63 for QCD jets and 0.75 for top jets, considering the 40 hardest constituents of each jet.}
Meanwhile the latter is shown in the third row of
Table~\ref{tab:aucs_baseline}, indicating OmniLearn AUCs of 0.9940 and 0.9714 for QCD and tops respectively. We conclude that the separability of GPT jets from original continuous \textsc{JetClass} is driven entirely by the binning of the data -- this alone takes the data off-manifold with respect to the original continuous samples. By contrast, GPT jets are quite close to the binned \textsc{JetClass} that they were trained on.

\subsection{Top vs.\ QCD separation with GPT}
\label{sec:gpt_top_qcd}

We next turn to an evaluation of how well the GPT model reproduces the separation of tops vs.\ QCD. We compare three classifiers in Figure~\ref{fig:jetclassref_optimalroc} (each classifier is evaluated on the samples from the generative process it's defined/trained with): 
\begin{enumerate}
\item ``GPT Optimal": The GPT model's likelihood ratio discriminator. Note that here it doesn't matter whether one uses the autoregressive likelihood for tokens $\log p_\theta(\t{\mathbf t}_{\mathbf{x}})$ or the bin-smeared likelihood in the continuous physical space $\log p_\theta^{cont}(\mathbf{x})$ -- the extra correction factors in (\ref{eq:logpcont}) cancel out.
\item ``JetClass OmniLearn": An OmniLearn classifier trained on \textsc{JetClass} jets.  
\item ``GPT OmniLearn": An OmniLearn classifier trained on samples drawn from the GPT model.  
\end{enumerate}
We see that while the OmniLearn classifiers trained on original \textsc{JetClass} and on GPT outputs agree, the NP-optimal classifier derived from the GPT model gives a much higher ROC curve. 
Evidently, the GPT tops vs.\ QCD jets are highly separable, and in a way that the OmniLearn classifier cannot detect. This reproduces the observation of Ref.~\cite{Geuskens:2024tfo}, and in our setup the separability is even more pronounced. %

However, unlike that work, we point out that there are two possible interpretations of this result: either 
\begin{enumerate}[label=(\alph*)]
\item 
The NP-optimal classifier between GPT tops and QCD 
is representative of the NP-optimal classifier between \textsc{JetClass} tops and QCD,
in which case there is a major gap between state-of-the-art jet taggers and the true ``fundamental limit" of jet tagging (the conclusion reached in \cite{Geuskens:2024tfo}), 
\end{enumerate}
or
\begin{enumerate}[label=(\alph*),start=2]
\item the GPT model introduces unphysical artifacts into \textsc{JetClass} that make the GPT tops much more separable from GPT QCD than their \textsc{JetClass} counterparts, and in a way that the classifier is somehow unable to learn. Then the fact that the state-of-the-art classifiers cannot reproduce the NP-optimal result on GPT-generated jets is unfortunate, but possibly beside the point if this result is due to unphysical artifacts.
\end{enumerate}

We next turn to another popular generative modeling framework for jets -- permutation-equivariant conditional flow matching -- and explore what it has to say about the separation of top vs.\ QCD jets.

\section{EPiC-FM: jets as particle clouds}
\label{sec:epicfm}

\subsection{Description of the model}
\label{sec:epicfm_description}

We represent jets as \emph{unordered} particle clouds with kinematic features
$\mathbf{x}=\{\mathbf{x}_i\}_{i=1}^{N}$, where each constituent carries $\mathbf{x}_i=( p_T,\Delta\eta,\Delta\phi)_i$. We restrict each jet to at most the 40 highest-$p_T$ constituents (i.e. $N_\text{max} = 40$), discarding softer particles beyond this cutoff. We use the conditional flow-matching (CFM) paradigm 
to train a continuous normalizing flow equipped with a time-dependent EPiC architecture \cite{Buhmann:2023zgc} that approximates the velocity field ${\bf u}_t$ of the ODE trajectories connecting source to target samples. The setup for our architecture closely follows the one in~\cite{Birk:2023efj}, with the main difference being that the only conditional model inputs that we consider are the time variable $t\in[0,1]$, the jet-type (QCD and top), and the number of constituents in each jet. For further details on the model and training see appendix~\ref{sec:epicfm_details}.

Once trained, we generate jets of a specified class by integrating the ODE
$\dot{\mathbf{x}}_t=\mathbf{u}_t^\theta(\mathbf{x}_t)$ from $t=0$ to $t=1$
with a forward Euler integrator with step size $\Delta t=3.33\times 10^{-3}$. We estimate the log-density of the generated jets by integrating the dynamics with backward Euler steps in the reverse time direction. Writing the total derivative as $d/dt=\partial_t+\mathbf{u}_t^\theta\!\cdot\nabla_{\mathbf{x}}$, the log-density evolves according to
\begin{align}
\frac{d}{dt}\log p_t(\mathbf{x}) \;=\; -\,\mathrm{Tr}\!\left[\frac{\partial \mathbf{u}_t^\theta}{\partial \mathbf{x}}\right],
\end{align}
which integrates to
\begin{align}
\log p_1(\mathbf{x}_1)
= \log p_0(\mathbf{x}_0)\;-\!\int_0^1\! dt\;
\mathrm{Tr}\!\left[\frac{\partial \mathbf{u}_t^\theta}{\partial \mathbf{x}}\right].
\end{align}
If the model is well trained, then this solution is expected to approximate the target log-density. In practice, we accumulate the Jacobian-trace term via automatic differentiation; numerical details and validation (fixed-point refinement, step-size scans) are provided in appendix~\ref{app:validation}.

As described in Sec.~\ref{subsec:gpt_desc}, tokenized jets are reconstructed in physical space by assigning each token to its corresponding $(\log p_T,\Delta\eta,\Delta\phi)$ bin and uniformly smearing it within that bin. Although EPiC-FM generates continuous jets rather than tokenized jets, its likelihood can still be evaluated on tokenized jets by mapping them into the continuous space through this bin-smearing procedure.

To approximate the log-likelihood of tokenized EPiC-FM generated jets, we apply the same procedure to the continuous EPiC-FM samples. Specifically, we first bin the constituent features of the generated jets and map each constituent to the center of its corresponding three-dimensional bin, producing the binned jet $\hat{\mathbf{x}}_1$. We then uniformly smear each constituent within its bin by adding noise $\boldsymbol{\epsilon}_k$, sampled independently for each constituent.

The log-likelihood is approximated by averaging over $M$ such smearing realizations,
\begin{align}
\log p_{bs}(\mathbf{x}_1)
= \log\left(\frac{1}{M}\sum_{k=1}^{M} p_1(\hat{\mathbf{x}}_1+\boldsymbol{\epsilon}_k)\right),
\end{align}
where $\hat{\mathbf{x}}_1$ denotes the binned EPiC-FM generated jet with constituent features mapped to the centers of the $(\log p_T,\Delta\eta,\Delta\phi)$ bins, and $\boldsymbol{\epsilon}_k$ represents the noise that uniformly smears each constituent within its bin. This estimator becomes exact in the limit $M \to \infty$. In our implementation we choose $M=100$.

\subsection{Baseline evaluation of the model}
\label{sec:validate_surrogate}

We again perform a 2-sample classifier test between EPiC-FM generated jets and \textsc{JetClass} jets.
We find AUCs of $0.6729$ for QCD and $0.7703$ for tops.  
These results indicate that EPiC-FM captures the main features of the simulation with moderate levels of mismodeling likely due to finite training data and finite model capacity. 

\subsection{Top vs.\ QCD separation with EPiC-FM}
\label{sec:epic_top_qcd}
Next we investigate whether EPiC-FM accurately models the separation of top vs.\ QCD jets. In Figure~\ref{fig:jetclassref_optimalroc} we compare the ROC curves of three top vs.\ QCD classifiers (each classifier is evaluated on the samples from the generative process it's defined/trained with):
\begin{enumerate}  
\item ``EPiC-FM Optimal": The NP-optimal EPiC-FM likelihood ratio discriminator 
\item ``JetClass OmniLearn": An OmniLearn classifier
trained on \textsc{JetClass} jets.
\item ``EPiC-FM OmniLearn": An OmniLearn classifier 
trained on samples from the EPiC-FM model.
\end{enumerate}
We see from Figure~\ref{fig:jetclassref_optimalroc} that all three ROC curves are fairly consistent with one another. We point out that there are two possible interpretations of these results: either
\begin{enumerate}[label=(\alph*)]
\item 
The NP-optimal classifier between EPiC-FM tops and QCD 
is representative of the NP-optimal classifier between \textsc{JetClass} tops and QCD,
in which case there is a small gap between state-of-the-art jet taggers and the true ``fundamental limit" of jet tagging, 
\end{enumerate}
or
\begin{enumerate}[label=(\alph*),start=2]
\item the EPiC-FM model is smearing out real physical features in \textsc{JetClass} which results in smaller separability between EPiC-FM tops and QCD.
\end{enumerate}

While we cannot totally rule out the possibility that \textsc{JetClass} has subtle features that neither EPiC-FM nor OmniLearn can detect, the concurrence of \textsc{JetClass} OmniLearn with both EPiC-FM ROC curves is highly suggestive that the NP-optimal \textsc{JetClass} ROC curve is also close by, and that the SOTA classifiers are close to the fundamental limit.

In any case,  the OmniLearn classifier is evidently powerful enough to learn the true separation of the EPiC-FM top and QCD samples, unlike what we saw for the GPT jets.\footnote{The difference in R50 values between EPiC-FM Optimal and EPiC-FM OmniLearn is within a factor of 2. This is consistent with recent gains in top tagging from improved architectures and more data \cite{Brehmer:2024yqw,Bhimji:2025isp}.} Together with the 2-sample classifier tests described in the previous subsection, this gives us confidence that EPiC-FM is not introducing unphysical artifacts into \textsc{JetClass} and thus constitutes a valid surrogate reference model.

\begin{figure}[t]
    \centering
    \includegraphics[width=\linewidth]{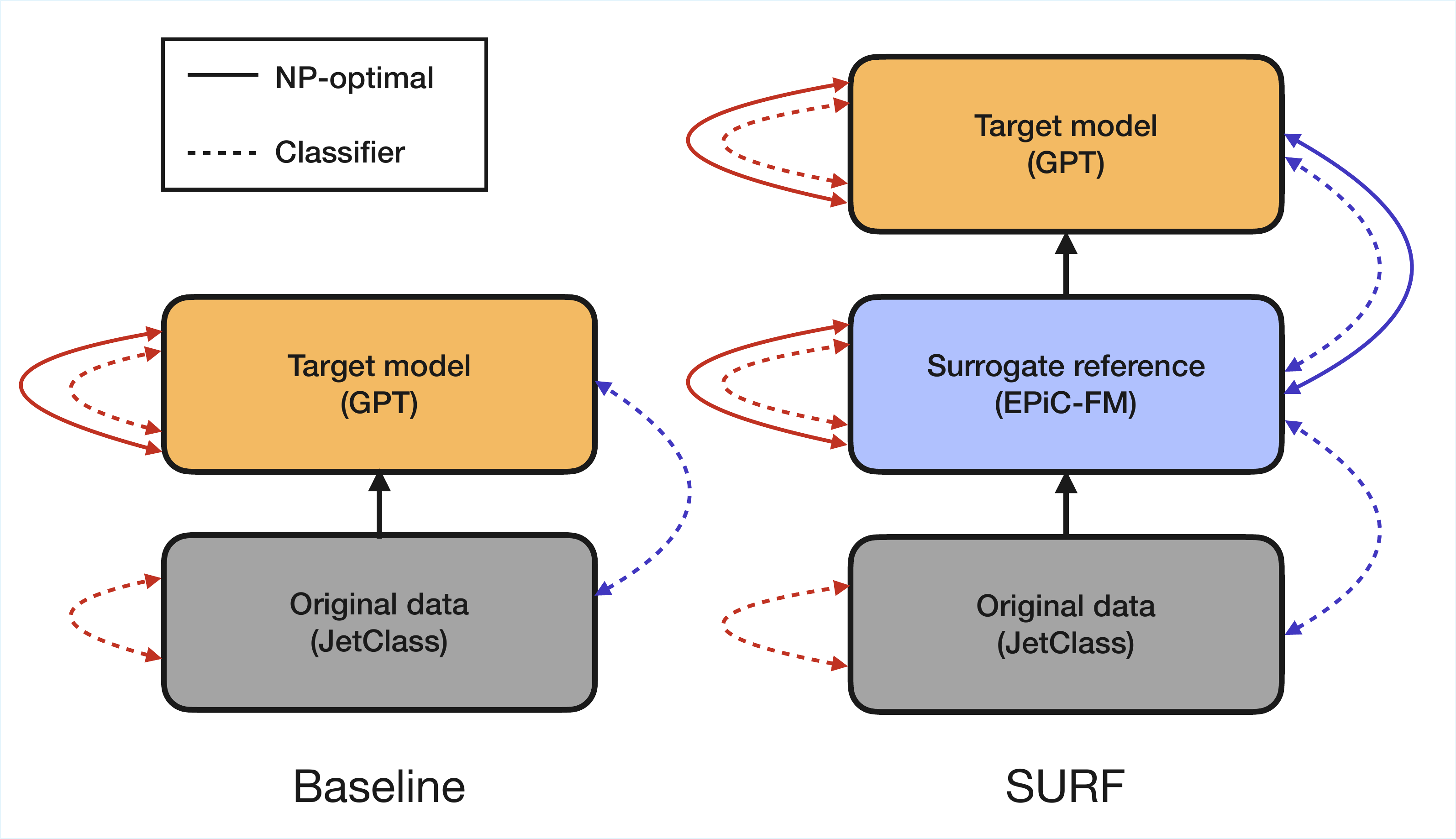}
    \caption{Illustration of the SURF method. The conventional baseline approach is shown on the left and the SURF approach is shown on the right. The red lines denote interclass (e.g. top vs QCD) tests and the blue lines denote 2-sample tests. The solid lines correspond to NP-optimal tests, while dashed lines correspond to trained classifier tests (e.g. OmniLearn). Here we see the NP-optimal tests enabled by the SURF method.}
    \label{fig:surf_diagram}
\end{figure}

\section{SURF Method}
\label{sec:surrogate_method}

We have two generative models that give different possibilities for the ``fundamental limit" of jet tagging -- GPT suggests a huge gap between the true NP-optimal ROC curve of tops vs.\ QCD and the performance of state-of-the-art classifiers, while EPiC-FM suggests almost no gap. Which is correct?

Here we propose a new method that provides a plausible answer to this puzzle: the {\it SUrrogate ReFerence (SURF) method}. As illustrated in Figure~\ref{fig:surf_diagram}, the core idea is to use a generative model with tractable likelihoods -- the \textit{surrogate reference model} -- to evaluate another target generative model whose likelihood is likewise tractable. We achieve this by first training the surrogate reference model on the original data, and then training the target model on samples generated by the surrogate reference model.

We have seen that EPiC-FM seems to be a valid surrogate for \textsc{JetClass}, in that it does  not seem to introduce any subtle artifacts into \textsc{JetClass} that OmniLearn cannot detect. So the SURF method in this context is to use samples from EPiC-FM as a new reference dataset to train GPT on. Then we will have access to all the classifiers as before, plus a fourth one: the true NP-optimal ROC curve for the reference dataset. This is what's lacking for \textsc{JetClass}, and is the value added by the SURF method.

\begin{table}
\centering
\renewcommand{\arraystretch}{2.5}
\begin{tabular}{|l|c|c|}
\hline
\textbf{Comparison} & \textbf{QCD} & \textbf{Top} \\ \hline
EPiC-FM vs GPT
  & \begin{tikzpicture}[baseline=(current bounding box.center)]
      \node[minimum width=2cm, minimum height=1.5cm] (box) {};
      \draw (box.south west) -- (box.north east);
      \node at ([shift={(-0.7cm,0.2cm)}]box.center) {0.9955};
      \node at ([shift={(0.4cm,-0.3cm)}]box.center) {1.000};
    \end{tikzpicture}
  & \begin{tikzpicture}[baseline=(current bounding box.center)]
      \node[minimum width=2cm, minimum height=1.5cm] (box) {};
      \draw (box.south west) -- (box.north east);
      \node at ([shift={(-0.7cm,0.2cm)}]box.center) {0.9746};
      \node at ([shift={(0.4cm,-0.3cm)}]box.center) {1.000};
    \end{tikzpicture} \\ \hline
EPiC-FM-bs vs GPT
  & \begin{tikzpicture}[baseline=(current bounding box.center)]
      \node[minimum width=2cm, minimum height=1.5cm] (box) {};
      \draw (box.south west) -- (box.north east);
      \node at ([shift={(-0.7cm,0.2cm)}]box.center) {0.5505};
      \node at ([shift={(0.4cm,-0.3cm)}]box.center) {0.5640};
    \end{tikzpicture}
  & \begin{tikzpicture}[baseline=(current bounding box.center)]
      \node[minimum width=2cm, minimum height=1.5cm] (box) {};
      \draw (box.south west) -- (box.north east);
      \node at ([shift={(-0.7cm,0.2cm)}]box.center) {0.5579};
      \node at ([shift={(0.4cm,-0.3cm)}]box.center) {0.6406};
    \end{tikzpicture} \\ \hline
EPiC-FM vs EPiC-FM-bs
  & \begin{tikzpicture}[baseline=(current bounding box.center)]
      \node[minimum width=2cm, minimum height=1.5cm] (box) {};
      \draw (box.south west) -- (box.north east);
      \node at ([shift={(-0.7cm,0.2cm)}]box.center) {0.9950};
      \node at ([shift={(0.4cm,-0.3cm)}]box.center) {0.9980};
    \end{tikzpicture}
  & \begin{tikzpicture}[baseline=(current bounding box.center)]
      \node[minimum width=2cm, minimum height=1.5cm] (box) {};
      \draw (box.south west) -- (box.north east);
      \node at ([shift={(-0.7cm,0.2cm)}]box.center) {0.9733};
      \node at ([shift={(0.4cm,-0.3cm)}]box.center) {0.9967};
    \end{tikzpicture} \\ \hline
\end{tabular}
\caption{
Same as Table~\ref{tab:aucs_baseline} but with \textsc{JetClass} replaced with the EPiC-FM surrogate reference. In each cell, the upper-left value corresponds to the OmniLearn classifier AUC,
and the lower-right value corresponds to the NP-optimal classifier AUC, which is now available thanks to the tractable likelihoods of the surrogate reference.
}
\label{tab:auc_surf}
\end{table}

As a first application of the SURF method, we perform the 2-sample classifier tests for top and QCD jets. These are summarized in the first and second rows of Table~\ref{tab:auc_surf} for continuous and bin-smeared jets respectively. We confirm using the EPiC-FM surrogate reference that the GPT-generated jets are well-separated from continuous jets, and that the OmniLearn classifier is able to pick this up. We also confirm 
that the GPT-generated jets are quite close to bin-smeared jets, so the OmniLearn classifier was not misleading on this front. As shown in the third row of Table \ref{tab:auc_surf}, we observe the same behavior previously identified for \textsc{JetClass}: bin-smeared EPiC-FM jets are nearly perfectly separable from their continuous counterparts.

\begin{figure*}
    \includegraphics[width=1.5\columnwidth]{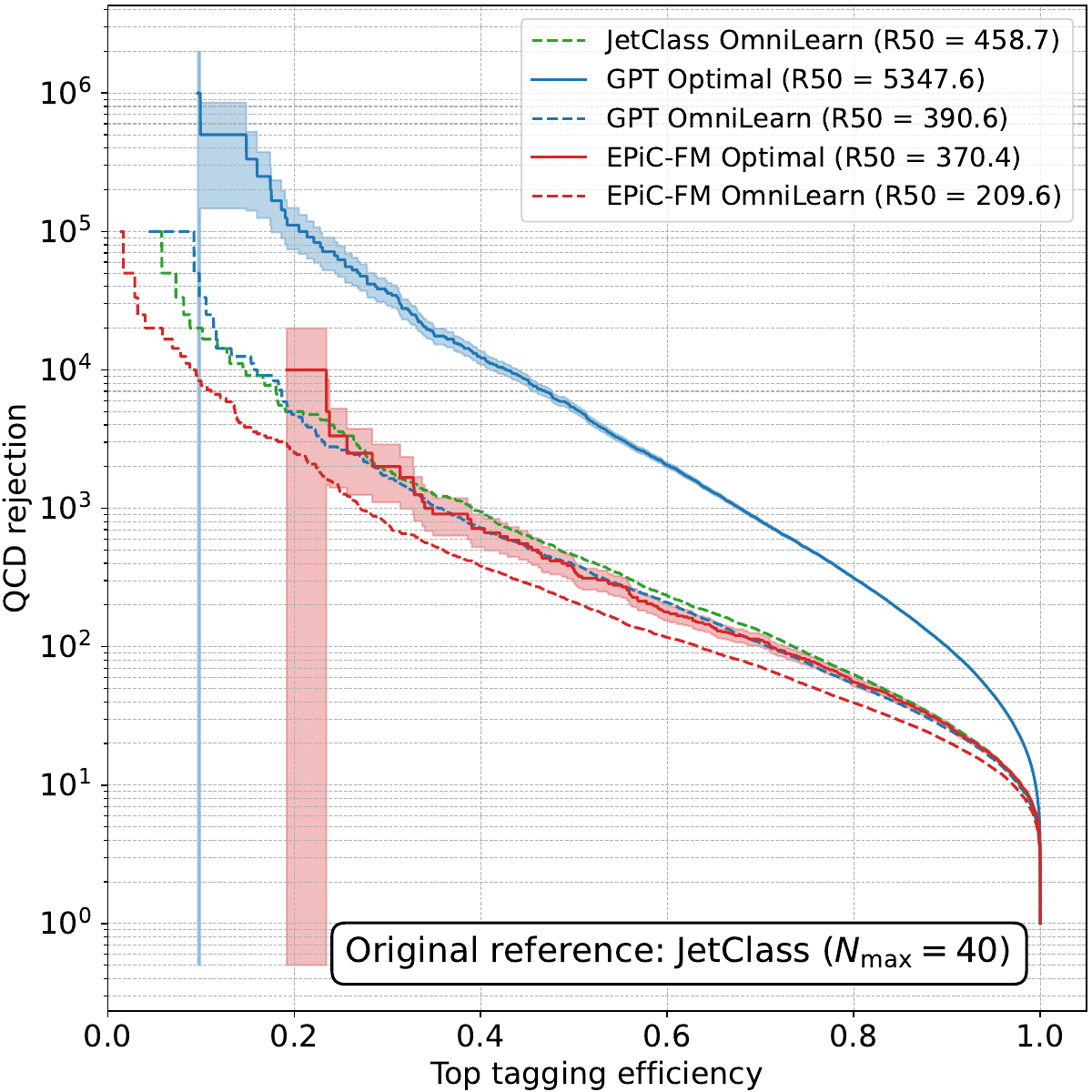}
    \caption{ROC curves for top vs.\ QCD jet classification derived from the \textbf{\textsc{JetClass}} dataset and the generative models trained on it. Note that, although physical jets can contain more than 40 constituents, in this study each jet is represented only by its 40 hardest constituents (i.e.~$N_\text{max} = 40$). Solid lines indicate the performance of the NP-optimal classifier obtained directly from the true log-likelihood ratio. Dashed lines correspond to classifiers trained with OmniLearn. The QCD rejection at a top tagging efficiency of 50\% (R50) is shown in parentheses for each curve. Shaded bands indicate the statistical uncertainty on the optimal ROC curves, estimated from binomial counting errors on the background sample and propagated to the QCD rejection axis.}
    \label{fig:jetclassref_optimalroc}
\end{figure*}

\begin{figure*}
    \includegraphics[width=1.5\columnwidth]{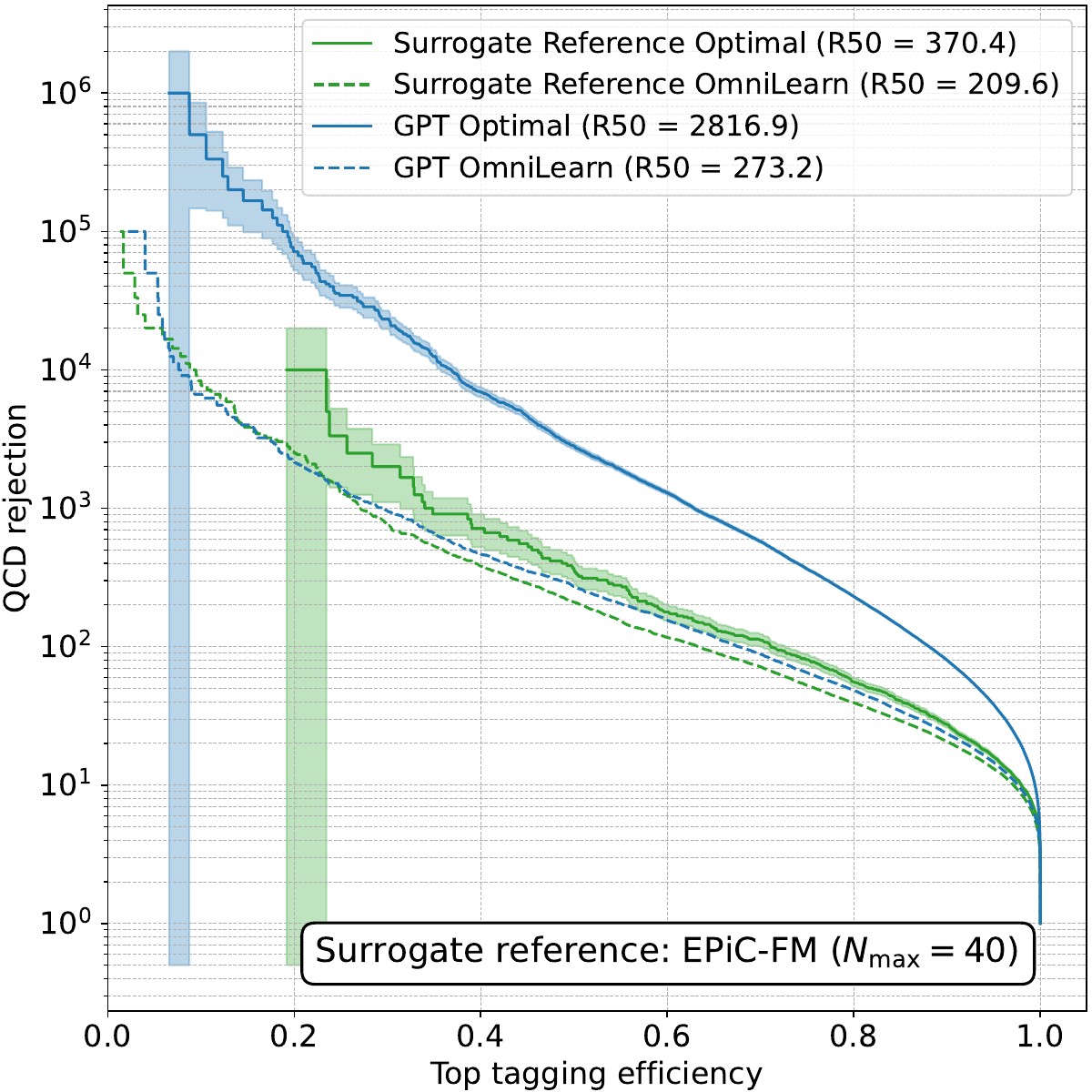}
    \caption{ROC curves for top vs.\ QCD jet classification derived from the \textbf{EPiC-FM surrogate reference} samples and the GPT models trained on them. Note that, although physical jets can contain more than 40 constituents, in this study each jet is represented only by its 40 hardest constituents (i.e.~$N_\text{max} = 40$).
    Solid lines indicate the performance of the NP-optimal classifier obtained directly from the true log-likelihood ratio. 
    Dashed lines correspond to classifiers trained with OmniLearn. 
    The QCD rejection at a top tagging efficiency of 50\% (R50) is shown in parentheses for each curve. 
    Shaded bands indicate the statistical uncertainty on the optimal ROC curves, estimated from binomial counting errors on the background sample and propagated to the QCD rejection axis.}
    \label{fig:epicfmref_optimalroc}
\end{figure*}

Next we turn to the main question of interest: whether these GPT models accurately represent the true degree of top vs.\ QCD separation or not.  The relevant ROC curves are shown in Figure~\ref{fig:epicfmref_optimalroc}. Here in addition to the analogues of the three classifiers considered previously -- 
GPT Optimal, 
Surrogate Reference OmniLearn and 
GPT OmniLearn, we are able to add a fourth ROC curve, Surrogate Reference Optimal, thanks to the tractable likelihoods of the EPiC-FM surrogate reference model.

We see that the GPT model vastly inflates the ground-truth ROC curve of the surrogate reference model! 
By contrast, the OmniLearn classifier curves follow the Surrogate Reference Optimal curve closely, showing that classifier-based discrimination remains consistent with the physically meaningful separation encoded in the EPiC-FM surrogate. This confirms that the inflated NP-optimal behavior originates from artifacts in the GPT likelihood landscape, not from deficiencies in classifier performance as was claimed in Ref.~\cite{Geuskens:2024tfo}.

As expected, we observe the same behavior -- both qualitatively and quantitatively -- as in the baseline GPT evaluation (Figure~\ref{fig:jetclassref_optimalroc}), where we found the NP-optimal ROC curve for GPT jets to be substantially 
larger than the OmniLearn classifiers trained on the reference data and on GPT samples. This cross-validation builds further confidence in our interpretation of the hugely inflated NP-optimal ROC curve from GPT tops and QCD.

\section{Investigating ROC Inflation in the GPT Model}
\label{sec:roc_inflation}

Since top and QCD jets are already extremely well separated (AUC $\geq 0.98$ with state-of-the-art taggers), the ROC curve of any approximate model becomes highly sensitive to even small mismodelings of the jet distributions. 

We illustrate this sensitivity using a 10D Gaussian toy model with background given by $\mathcal{N}(\mathbf{0}, I_{10})$ and signal given by $\mathcal{N}(\mathbf{1}, I_{10})$. To emulate mismodeling, we introduce shifted versions of the background and signal, described by Gaussians with slightly displaced means: $\mathcal{N}\!\left(-0.1\,\mathbf{1}_{10}, I_{10}\right)$ and $\mathcal{N}\!\left(1.1\,\mathbf{1}_{10}, I_{10}\right)$. 

\begin{figure}[t]
    \centering
    \includegraphics[width=0.9\linewidth]{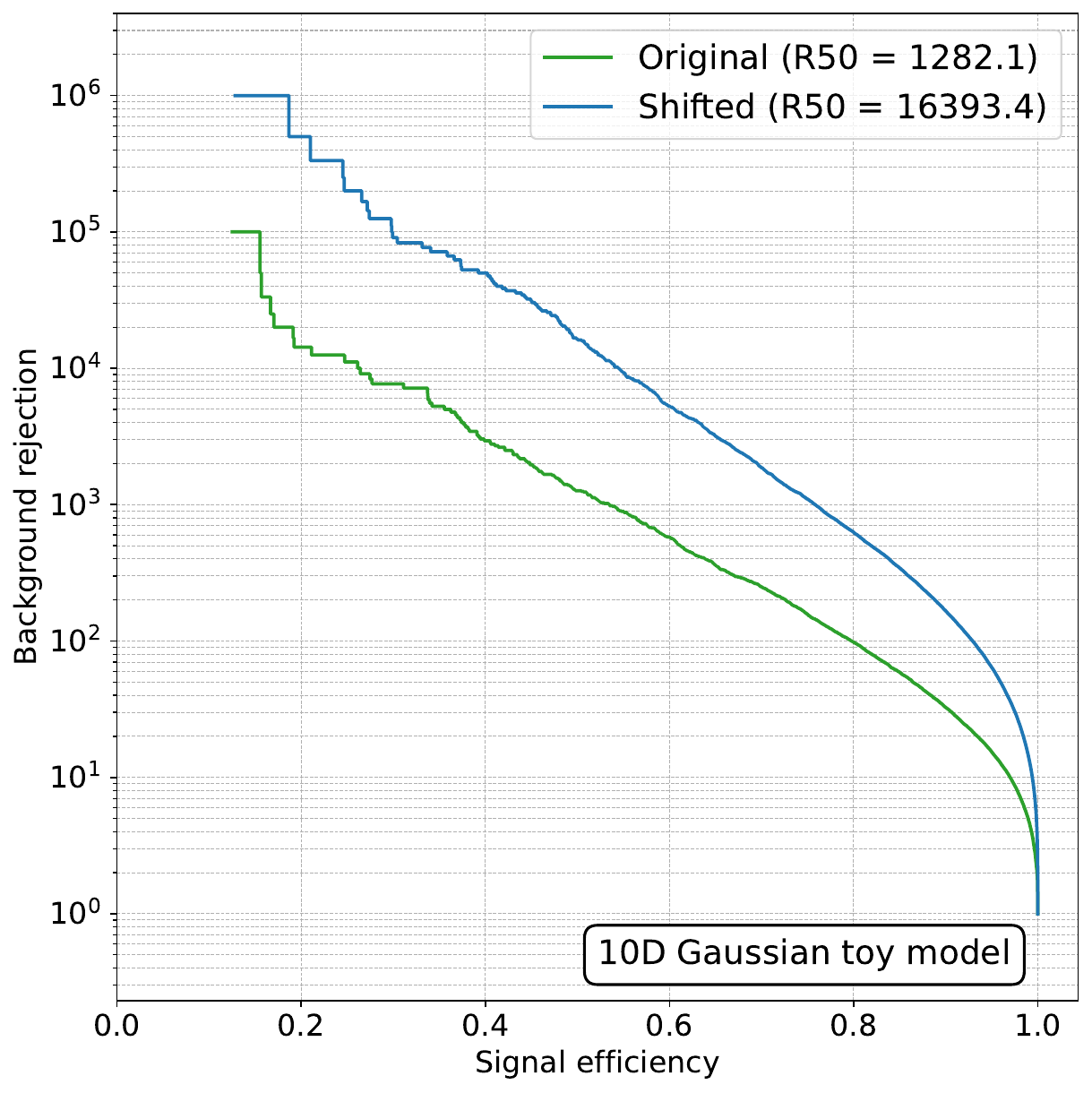}
    \caption{ROC curves from the 10-dimensional Gaussian toy model. 
    Although the shifted signal and background distributions differ only slightly from their original counterparts, 
    their mutual ROC curve appears highly inflated. 
    Here, ``Original" denote the original signal and background, 
    while ``Shifted" indicate the slightly displaced versions. The background rejection at a signal efficiency of 50\% (R50) is shown in parentheses for each curve.
    This demonstrates how small mismodelings of well-separated classes can artificially exaggerate separation power.}
    \label{fig:inflated_roc}
\end{figure}

While the AUCs between the original and shifted distributions are around $0.6$, the ROC curve between the shifted signal and background appears highly inflated, as shown in Figure~\ref{fig:inflated_roc}. Because the original signal and background already exhibit minimal overlap, even a small shift in the likelihood ratios can have an outsized impact on the ROC curve. This example highlights how minor deviations in otherwise well-separated classes can artificially enhance apparent separation power.

Next we turn to the question of what kind of mismodeling could be causing the inflated ROC curve in the GPT-generated jets.

\subsection{Bin-smearing?}

The fact that the GPT-generated jets are attempting to match bin-smeared jets, and the bin-smeared jets are off-manifold compared to continuous jets, could be a significant source of mismodeling. 

\begin{figure}
    \centering
    \includegraphics[width=0.9\linewidth]{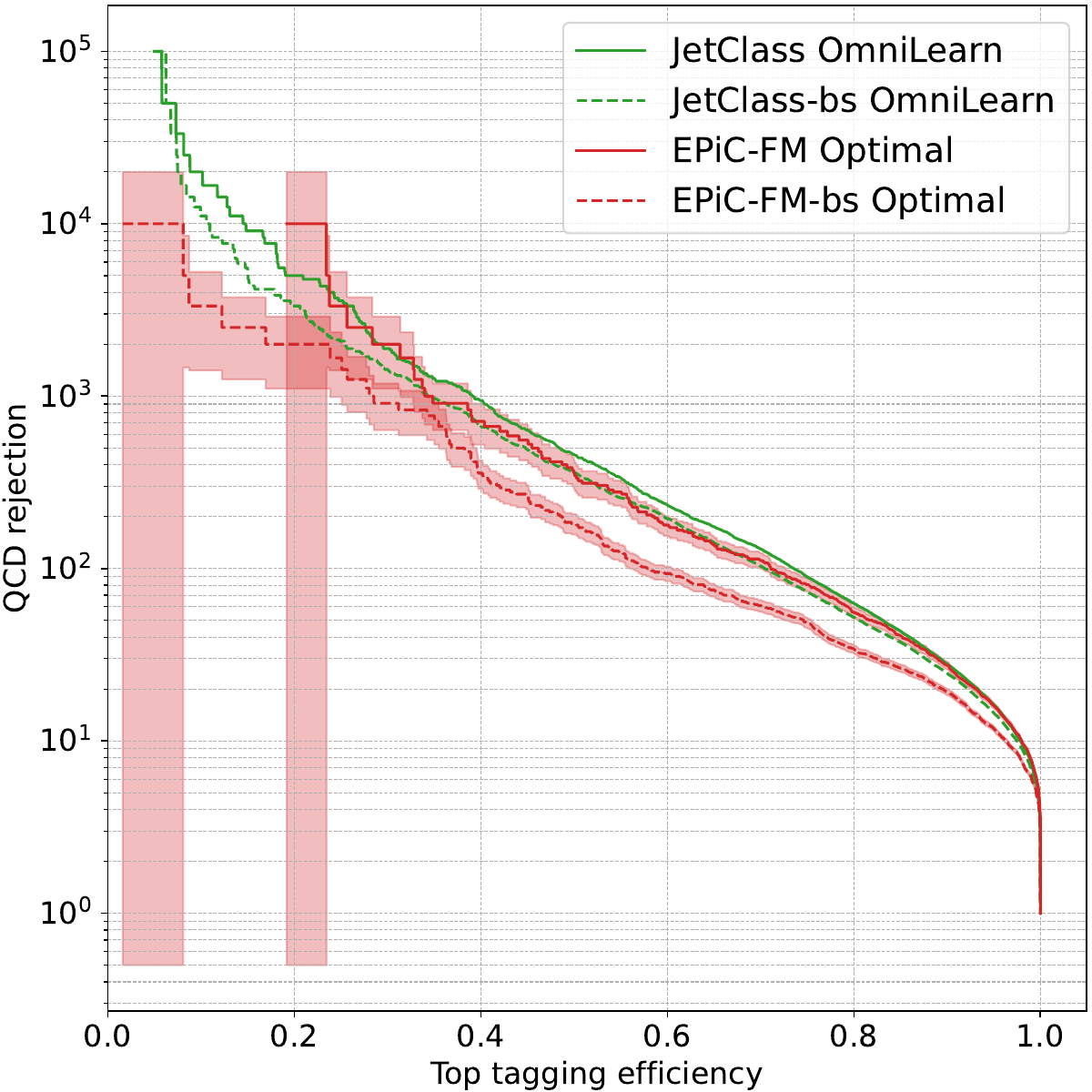}
    \caption{Comparison of top vs.\ QCD discrimination for both EPiC-FM surrogate reference jets and \textsc{JetClass} jets. The EPiC-FM curves, shown in red (continuous) and red dashed (bin-smeared, labeled ``bs"), represent the true NP optimal. Shaded bands indicate the statistical uncertainty on the optimal ROC curves, estimated from binomial counting errors on the background sample and propagated to the QCD rejection axis. For \textsc{JetClass}, where the true likelihood is not available, the solid and dashed green curves show the corresponding OmniLearn classifiers trained on continuous and bin-smeared (``bs") \textsc{JetClass} samples, respectively. Across both EPiC-FM and \textsc{JetClass} jets, bin-smearing leads to slightly reduced separability between top and QCD jets.}

    \label{fig:binsmear_hypothesis}
\end{figure}

However, Figure~\ref{fig:binsmear_hypothesis} shows that this is not the issue. Using the EPiC-FM model as the reference, we can compute the true likelihood of both continuous and bin-smeared jets, so we can obtain the true ROC curve between continuous tops vs.\ QCD as well as bin-smeared tops vs.\ QCD. We see that if anything, the bin-smeared jets are {\it less} separable than their original continuous counterparts. For \textsc{JetClass}, where the true likelihood is not available and we instead rely on an OmniLearn classifier, we observe the same qualitative behavior that bin-smearing reduces separability. So the binning of the jets, despite being a major deviation from the continuous jets, cannot be the cause of the inflated top vs.\ QCD ROC curve. 

\subsection{Overfitting}

\begin{figure*}[t]
  \subfloat[EPiC-FM train and validation loss curves.\label{fig:fm_loss}]{
    \includegraphics[width=0.47\textwidth]{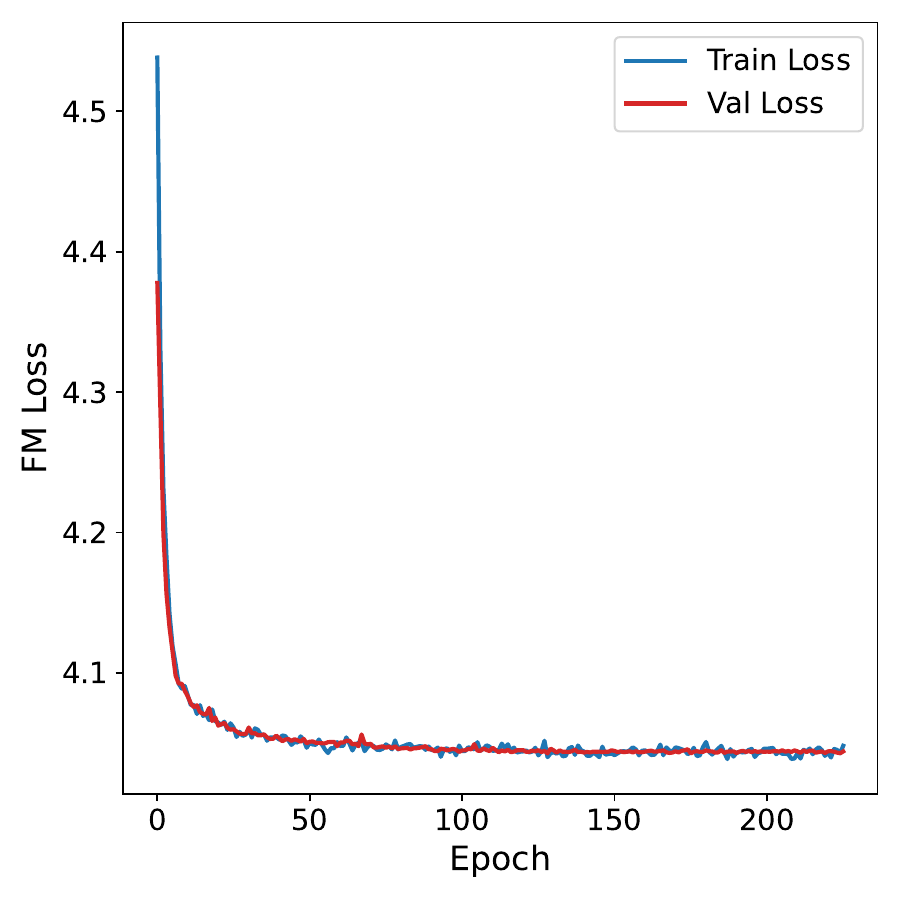}
  }
  \hfill 
  \subfloat[GPT (tops) train and validation loss curves.\label{fig:gpt_loss}]{
    \includegraphics[width=0.478\textwidth]{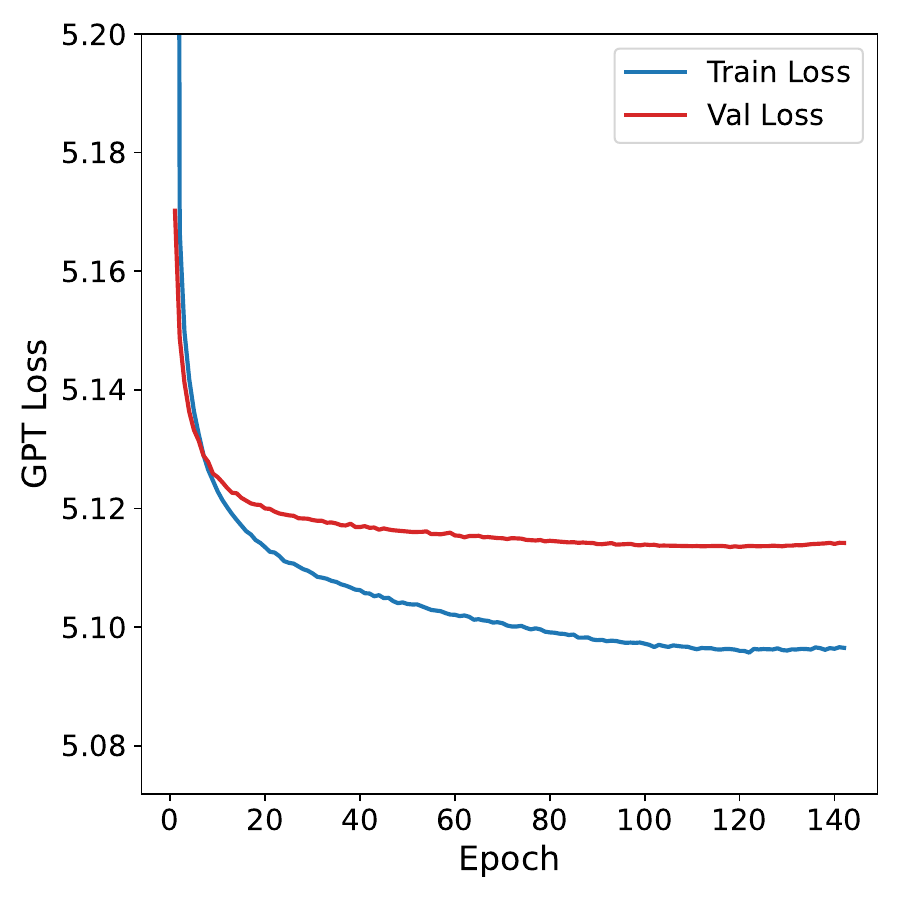}
  }

  \caption{Training vs.\ validation loss curves for EPiC-FM and GPT.  
  The GPT plot is shown for tops, with qualitatively similar behavior observed for QCD.  
  GPT overfits almost immediately (diverging train/val losses), while EPiC-FM maintains closer alignment.}
  \label{fig:loss_curves}
\end{figure*}


\begin{figure}[t]
    \centering
    \includegraphics[width=0.9\linewidth]{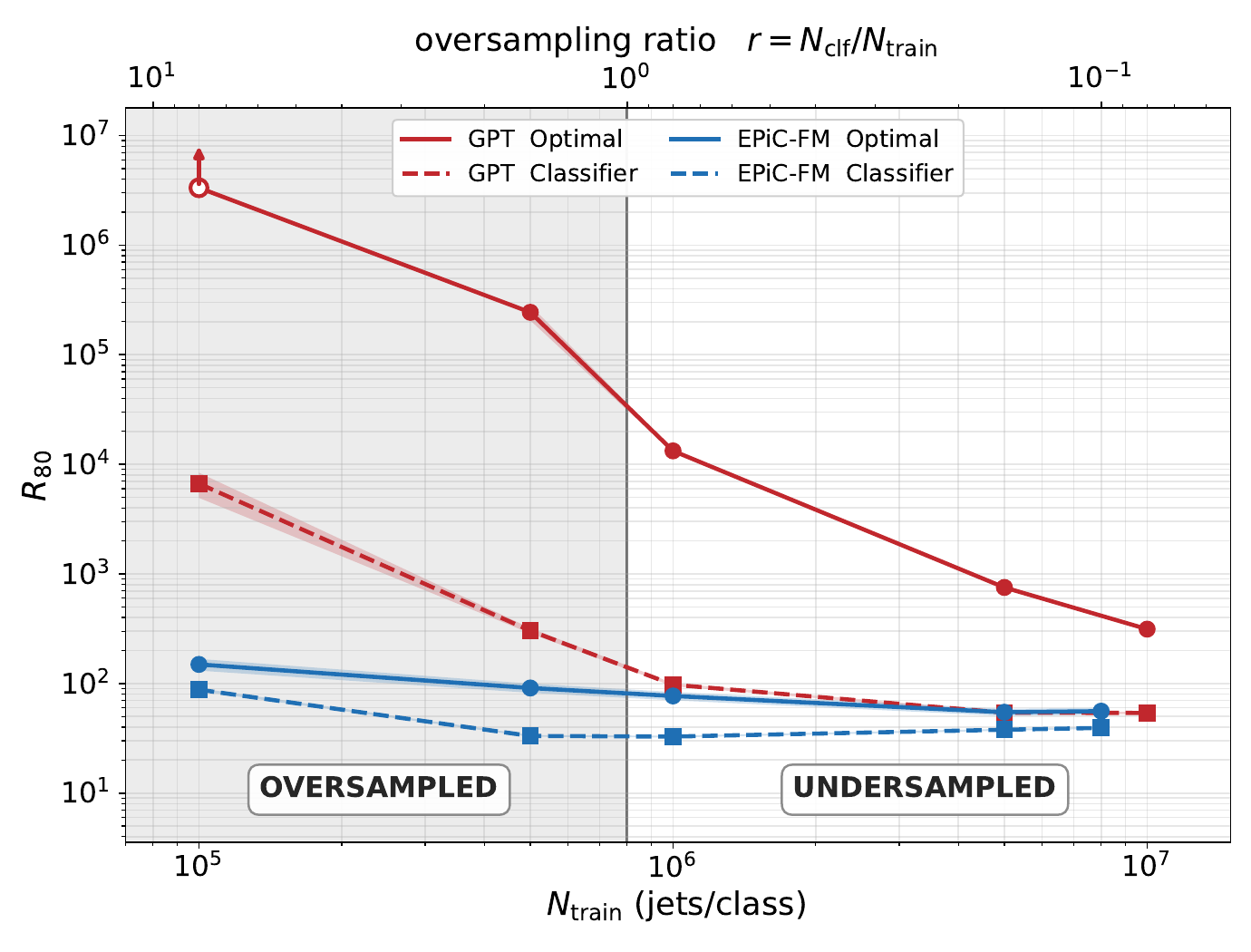}
    \caption{QCD rejection at $80\%$ top efficiency ($R_{80}$) for GPT and EPiC-FM models trained on increasing numbers of jets per class, $N_\mathrm{train}$. Solid lines show the NP-optimal values, while dashed lines show the corresponding OmniLearn classifier values. The OmniLearn classifiers are trained on a fixed $N_\mathrm{clf} = 8\times10^5$ generated jets per class, independent of $N_\mathrm{train}$. The top axis gives the oversampling ratio $r = N_\mathrm{clf}/N_\mathrm{train}$, and the shaded region marks $r > 1$, where the classifier is trained on more generated jets than the generator itself saw. At the smallest training size, no GPT background jets survive the cut, so its optimal $R_{80}$ is shown as a $95\%$ CL lower limit.}

  \label{fig:vary_overfitting}
\end{figure}

Finally, we consider another potential source of the mismodeling that affects GPT-generated jets but not EPiC-FM-generated jets: overfitting. By overfitting, we refer to the situation where the model achieves higher performance on the training set than on the validation set. In our case, an overfitted model produces jets that more closely resemble those in the training dataset than those in the validation dataset. Shown in Figure~\ref{fig:loss_curves} are comparisons of the train vs.\ validation loss curves for GPT and EPiC-FM models. We see that the GPT jets overfit almost immediately (increasing gap between train vs.\ val losses), despite  their val loss continuing to decrease slowly. Overfitting in GPT-style models is a known phenomenon, discussed already in the original GPT-2 paper~\cite{Radford2019LanguageMA}. Meanwhile the EPiC-FM loss curves are in perfect agreement between train vs.\ validation. Hence, we do not find any evidence for the EPiC-FM model overfitting.\footnote{ Although the EPiC-FM loss is not equivalent to the log-likelihood, we checked that the EPiC-FM model with the best validation loss gives consistent log-likelihoods on the train and validation set.} More broadly, flow-based models such as diffusion and flow-matching have often been observed to generalize well in machine learning applications, though the underlying reasons remain an active area of investigation (see e.g.,\, Refs.~\cite{kadkhodaie2023generalization,vastola2025generalization,bertrand2025closed}).


Finally, we directly examine the effects of model overfitting on the optimal likelihood ratio performance. We train separate GPT and EPiC-FM models at fixed capacity on subsets of \textsc{JetClass} ranging from $10^5$ to $10^7$ jets for tops and QCD. As the training set size decreases -- particularly when the number of examples becomes small relative to the number of model parameters and in the absence of strong regularization -- GPT overfitting becomes more severe. Indeed, as shown in Figure~\ref{fig:vary_overfitting}, the GPT models trained on smaller datasets exhibit a strongly inflated NP-optimal $R_{80}$, spanning several orders of magnitude, indicating that overfitting amplifies the apparent separation. The EPiC-FM models, by contrast, show only mild inflation, consistent with the loss curve comparison in Fig.~\ref{fig:loss_curves} that finds no evidence of EPiC-FM overfitting.

The overfitting hypothesis could also potentially explain why state-of-the-art  classifiers cannot detect the inflated ROC curve for GPT jets. Overfitting in generative models means that the model is too close to the training data, starting to memorize the training data and become a lookup table, i.e.\ the model is reducing to a sum of (smeared out) delta functions centered around the training data. Compared to the smooth distribution describing the reference data, a sum of smeared out delta functions is a high-frequency deformation. In the GPT case, this shows up as memorization of specific token co-occurrence patterns within the training jets, in the extreme, near-copying of jet (sub)sequences. Such behavior can distort multi-particle (higher-order) correlations, manifesting as ‘high-frequency’ artifacts in the data manifold. Neural networks are well-known to have difficulty learning high frequency modes; see Refs.~\cite{rahaman2019spectral,ronen2019convergence} for discussions and studies of this. We speculate that this could be the ultimate reason why the state-of-the-art classifiers cannot detect the inflated GPT ROC curves. 



To investigate this explanation, we trained OmniLearn classifiers on the jets generated by the GPT and EPiC-FM models used in Fig.~\ref{fig:vary_overfitting}, using $8 \times 10^5$ jets per class for training, with a further $10^5$ per class each for validation and testing. We observe the following behavior. For GPT, the NP-optimal $R_{80}$ inflates monotonically as $N_\mathrm{train}$ decreases, by orders of magnitude across the full range. The classifier $R_{80}$, by contrast, remains roughly constant at large $N_\mathrm{train}$, indicating that the dominant features of the learned distribution, which are all the classifier can access in this regime, are stable. It inflates only once $N_\mathrm{train}$ falls below the classifier training size, and even then far less than the NP-optimal value, so the gap between the two grows sharply. For EPiC-FM, both the NP-optimal and classifier $R_{80}$ inflate only mildly and stay close to one another throughout.

This behavior is consistent with our interpretation. When a generative model overfits on limited training data, it introduces fine-grained structure in the learned density. Such structure is present in the likelihood ratio, and therefore in the NP-optimal ROC curve, but a trained classifier can only reliably exploit it if it is sufficiently sampled. The classifier is trained on $8 \times 10^5$ generated jets per class, so for generators trained on fewer jets we are in an oversampling regime, where these fine-grained features are repeatedly sampled and can be partially learned; the classifier $R_{80}$ inflates accordingly, only once this regime is entered. Taken together with the loss curve comparison, the much weaker inflation of the EPiC-FM curves suggests that the continuous flow model overfits far less than the discrete GPT model, if at all. We caution, however, that the absence of detectable overfitting does not establish that EPiC-FM does not overfit, only that we find no evidence for it with the diagnostics used here.

\section{Conclusion}
\label{sec:conclusion}

We revisited recent claims that autoregressive GPT-style jet generators expose a large performance gap between ML-based jet taggers and the NP optimal ``fundamental limit'' of top vs.\ QCD discrimination~\cite{Geuskens:2024tfo}. Using our own independently trained GPT-based jet generator -- with an improved architecture from that used in the original study -- we reproduced their reported behavior. However, when comparing with the tractable EPiC-FM flow-matching model, which provides explicit likelihoods for unordered particle clouds, we find no evidence for such an inflated limit.

To resolve the apparent tension between the EPiC-FM and GPT results, we introduced the SURF method, which enables exact NP tests even when the real data likelihood is intractable. By training the target GPT model on samples from a tractable EPiC-FM surrogate, SURF allows direct NP comparisons between the surrogate and the target. We find that the GPT likelihood ratios dramatically overstate the true top vs.\ QCD separation encoded in the EPiC-FM surrogate. The ``fundamental limit'' ROCs obtained from GPT models therefore reflect artifacts of the generative process rather than performance limitations of the jet taggers.

We speculate that the inflated GPT ROC curve could be due to overfitting, and more generally high frequency artifacts introduced by the GPT model. Such high-frequency modes can be exploited by exact NP tests but are difficult for neural-network classifiers to learn, given their well-known low-frequency spectral bias~\cite{rahaman2019spectral,ronen2019convergence}. In future work it would be interesting to explore these high frequency artifacts in more detail. Existing work in the ML literature (e.g.,~\cite{zhang2023diffusion}) suggests potential methods for making such artifacts more explicit. It would also be interesting to develop new classifier architectures that can detect these high frequency artifacts. This would lead to more sensitive classifier metric tests. 

Furthermore, our study shows that GPT models trained on tokenized jets are faithful only to their discretized references, while being nearly perfectly separable from the original continuous jets. These results indicate that discretization can introduce significant off-manifold effects, which may limit the fidelity of generative models based on binning and tokenization if not carefully mitigated. Alternative discretization schemes could potentially reduce the degree of separability from the continuous reference. For example, Refs.~\cite{Golling:2024abg,Birk:2024knn} use VQ-VAEs to tokenize jet constituents. Exploring such strategies, and their impact on preserving on-manifold fidelity, would be an interesting direction for future study.

Overall, the consistent picture between \textsc{JetClass} and the EPiC-FM surrogate reference, particularly the concurrence of three ROC curves -- classifiers trained on samples from \textsc{JetClass} and EPiC-FM samples, and the NP-optimal likelihood ratio from EPiC-FM -- point to the fundamental limit of top tagging being not far from the performance of state-of-the-art classifiers. However, we cannot rule out the possibility that there remains a gap between the fundamental limit of top tagging and SOTA classifiers -- the EPiC-FM surrogate and OmniLearn may both miss subtle features in \textsc{JetClass} that make tops and QCD more separable than they seem. 

We can think of several avenues for further probing the fundamental limit of jet tagging. For example, improving the surrogate reference should be a high priority. Our EPiC-FM surrogate was far from perfect -- our classifier tests showed that they are partially separable from the original continuous jets (AUC$\approx$0.7). It would be worthwhile to repeat this study with a more expressive surrogate reference (for example replacing EPiC with a transformer) that is closer to the \textsc{JetClass} data to confirm the conclusions reached here. It would also be good to repeat this study with a more faithful target generative model that does not inflate the true ROC curve of the surrogate reference. Taken together (a more faithful surrogate and target), this could start to reveal a gap between SOTA classifiers and the fundamental limit. 

Recently some studies were initiated into the true theoretical limits on top tagging \cite{Larkoski:2024hfe}. It would be interesting to bring the empirical, data-driven approach explored here into contact with the theory-driven approach and elucidate further the true fundamental limit of jet tagging.

We focused on  just the momentum four-vectors of jet constituents in this work. For many jet tagging tasks, including top tagging, additional features are highly informative, such as track displacements and particle ID. These features are also included in \textsc{JetClass}, and it would be interesting to repeat this study for the full set of \textsc{JetClass} features to obtain an even fuller picture of the fundamental limit of jet tagging.

Finally, the SURF method provides a general route to perform exact NP tests using a tractable surrogate reference. Extending this approach to other classes of generative models, datasets, and applications -- both within and beyond high-energy physics -- would enable more systematic analysis of generative models.

\section*{Code availability}
The code for the GPT model and the EPiC-FM model can be found at \href{https://github.com/dfaroughy/JetSequences}{github.com/dfaroughy/JetSequences} and \href{https://github.com/Ian-Pang/jet_cfm}{github.com/Ian-Pang/jet\_cfm}, respectively.

\section*{Acknowledgements}
IP, DAF, DS and RD are supported by DOE grant DOE-SC0010008. RD is also supported by Deutsche Forschungsgemeinschaft (DFG, German Research Foundation) under grant 396021762 – TRR 257 Particle Physics Phenomenology after the Higgs Discovery. GK is supported by the DFG under the German Excellence Initiative -- EXC 2121  Quantum Universe – 390833306. This research used resources of the National Energy Research Scientific Computing Center, a DOE Office of Science User Facility supported by the Office of Science of the U.S. Department of Energy under Contract No. DE-AC02-05CH11231 using NERSC award HEP-ERCAP0027491.

\appendix

\section{Details of the EPiC-FM model}

\subsection{Architecture and training}
\label{sec:epicfm_details}

The EPiC network is composed of EPiC layers~\cite{Buhmann:2023pmh} -- a permutation-equivariant variant of Deep Sets~\cite{zaheer2017deep}, also known in high-energy physics as particle flow networks (PFNs)~\cite{Komiske:2018cqr}. These layers couple per-particle embeddings to a learnable global context, enabling the model to capture particle correlations while exactly preserving the permutation symmetry of the constituents~\cite{Buhmann:2023zgc}. The setup for our architecture closely follows the one in~\cite{Birk:2023efj}, with the main difference being that the only conditional model inputs that we consider are the time variable $t\in[0,1]$, the jet type (QCD and tops), and the number of constituents in each jet. Time is embedded into a vector space $\mathbb{R}^{16}$ using a sinusoidal embedding network and the jet-type labels are embedded into $\mathbb{R}^8$ using a linear lookup table. We stack $12$ EPiC-Layers with a local hidden dimension of $h_{\rm loc}=300$ and a global hidden dimension $h_{\rm glob}=32$. The variable number of particles per jet is handled via binary masking. During training the multiplicity is given by the true number of constituents in each jet, while during generation it is sampled from an empirical distribution computed from the \textsc{JetClass} dataset.

During training, we construct source–target pairs by sampling the source from a three-dimensional Gaussian and randomly pairing it with a target jet (QCD or top) from the dataset. We use 10 million top jets and 10 million QCD jets, split 80/20 into train/validation. 

Optimization uses {\tt Adam}~\cite{kingma2014adam} with an initial learning rate of $10^{-4}$ and a reduce-on-plateau schedule triggered by the validation loss with a patience of 10 epochs. We train for a total of 1000 epochs with a patience window of $50$ epochs. We select the best checkpoint with the lowest validation loss.

\subsection{Validation of log-likelihood computation}
\label{app:validation}

With a trained FM model $\*u^\theta$, we can calculate the log-likelihood iteratively by taking the following inference step
\begin{align}
    \log p_{t-\Delta t}(\*x_{t-\Delta t}) =\log p_{t}(\*x_{t})-\Delta t\, {\rm Tr}( {\tt Jac}[\*u^\theta_{t-\Delta t}(\*x_t)])\,,\nonumber
\end{align}
at every time step $\Delta t$. We compute the Jacobian matrix {\tt Jac[]} of the neural network in pytorch via {\tt torch.func.jacrev}. We verified that using {\tt torch.autograd.functional.jacobian} yields identical results.

A natural concern is whether flow matching models yield \emph{reliable} likelihood \emph{ratios}, especially in light of claims about the precision of likelihood computation in related models~\cite{Geuskens:2024tfo}. Constant offsets in log-likelihoods cancel in this ratio and therefore do not affect ROC curves. Operationally, we require numerical stability of \emph{differences} in log-likelihoods accumulated along the FM trajectory. We assessed this with the following two tests. 

\subsubsection{Consistency of Forward and Backward Euler Integration}

As mentioned in Sec.~\ref{sec:epicfm_description}, the forward Euler step is used for generation and the backward step is used for log-likelihood computation. Therefore, it is vital that the backward step is 
as close as possible to an
exact inverse of the forward step. This  guarantees consistent trajectories and preserves the theoretical assumption that the underlying flow is perfectly reversible. Any mismatch between the two steps violates this assumption and might lead to incorrect log-likelihood computation.

Unlike for regular normalizing flows, consistency of forward and backward steps is not guaranteed for continuous normalizing flows. The reason can be seen by the following:
\begin{align}
    \text{Forward step:}&\quad \mathbf{x}_{t+\Delta t} = \mathbf{x}_t + \Delta t \, \mathbf{u}_t^\theta(\mathbf{x}_t)\,, \label{eq:forward}\\
     \text{Backward step:}&\quad \mathbf{x}_{t}' = \mathbf{x}_{t+\Delta t} - \Delta t \, \mathbf{u}_t^\theta(\mathbf{x}_{t+\Delta t})\,. \label{eq:backward}
\end{align}
A backward update  (Eq.~\ref{eq:backward}) generally fails to recover the original point, since
\[
\mathbf{x}'_t - \mathbf{x}_t
= \Delta t\bigl(\mathbf{u}_t^\theta(\mathbf{x}_{t+\Delta t}) - \mathbf{u}_t^\theta(\mathbf{x}_t)\bigr)
\neq 0
\]
This means that the forward and backward updates are not exact inverses of each other; they only approach invertibility in the limit of infinitesimally small step sizes, $\Delta t \rightarrow 0$.

We observe that the backward step would be an exact inverse of the forward step if, in Eq.~\ref{eq:backward}, the vector field $\mathbf{u}_t$ is evaluated at $\mathbf{x}_t$ rather than at $\mathbf{x}_{t+\Delta t}$. However, this is precisely what is unavailable during the computation of the backward step, indeed this is meant to be the {\it output} of the backward step. 

\begin{figure*}[ht]
  \centering
  \makebox[\textwidth][c]{%
    \subfloat[Error suppression]{%
      \includegraphics[width=0.4\textwidth]{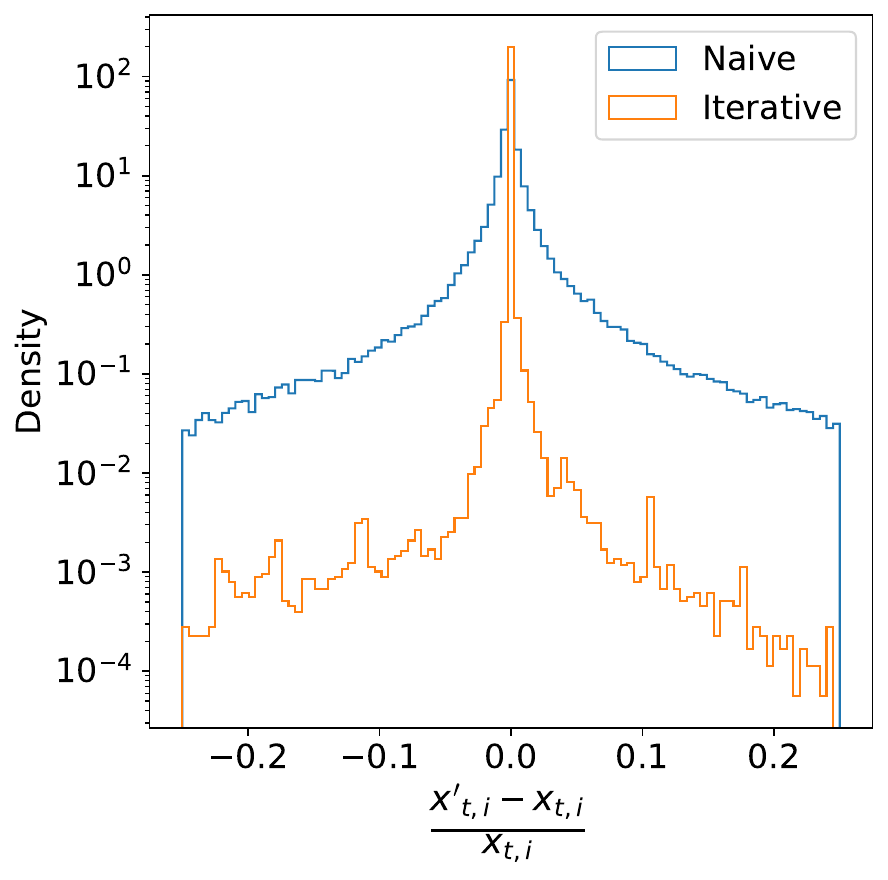}%
      \label{fig:iterative_combined_a}%
    }%
    \hspace{0.05\textwidth}%
    \subfloat[Classifier robustness]{%
      \raisebox{2.2ex}{%
        \includegraphics[width=0.4\textwidth]{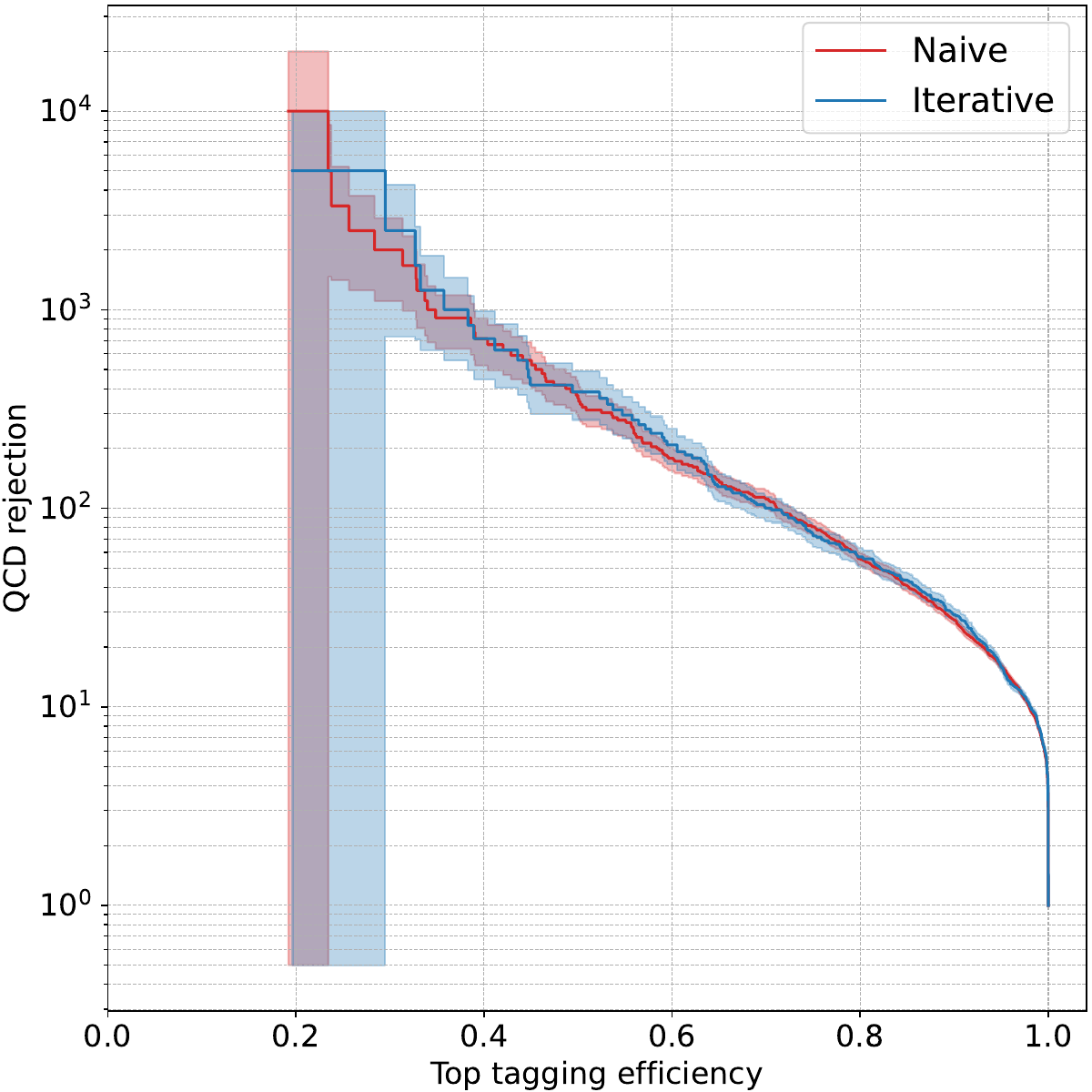}%
      }%
      \label{fig:iterative_combined_b}%
    }%
  }
  \caption{Effect of fixed‑point refinement in the backward update.  
    “Naive” denotes the backward update computed without any fixed‑point iterations;  
    “Iterative” denotes the backward update performed with five fixed‑point refinement iterations.  
    (a) Component‑wise relative error \(\tfrac{x'_{t,i}-x_{t,i}}{x_{t,i}}\) is sharply peaked around zero when using fixed‑point iterations. Here we show results for a few top jets; similar behavior is observed across larger jet samples, including QCD jets.  
    (b) The ROC curves built from FM‑based log‑likelihood ratios with and without refinement nearly coincide, showing negligible impact on classifier performance. Shaded bands indicate the statistical uncertainty on the optimal ROC curves, estimated from binomial counting errors on the background sample and propagated to the QCD rejection axis.}
  \label{fig:iterative_combined}
\end{figure*}

The problem is solved if we view the backward step as an {\it implicit equation} for $\mathbf{x}_t$:
\begin{equation}
    \mathbf{x}_t = \mathbf{x}_{t+\Delta t} - \Delta t\,\mathbf{u}_t^\theta(\mathbf{x}_t),
\end{equation}
This can be solved for $\mathbf{x}_t$ via fixed‐point iteration.

Figure~\ref{fig:iterative_combined_a} displays the distribution of the component‑wise relative error \[
\frac{x'_{t,i} - x_{t,i}}{x_{t,i}},
\]
where $i$ indexes all components of the flattened jet (i.e., both constituent and feature). Results are shown for backward updates computed with and without fixed-point refinement over 300 integration time steps. When employing the iterative backward update, we performed five fixed‑point refinement iterations. The iterative method yields a sharply peaked distribution around zero and strongly suppresses the heavy tails observed in the naive backward update. 

Nevertheless, we show in Figure~\ref{fig:iterative_combined_b} that the error has minimal impact on approximate NP classifier performance computed based on log-likelihood values obtained with and without the iterative procedure.\footnote{In the iterative case, error bands are wider, as computational constraints limited the evaluation sample size used to construct the ROC curve.} This study demonstrates the robustness of the FM-based log-likelihood ratio computation to errors accumulated in the backward update. For the main results presented in the paper, we employed the naive backward update to efficiently generate a large training dataset for the jet taggers at reasonable computational costs.

\begin{figure}
    \centering
    \includegraphics[width=0.9 \columnwidth]{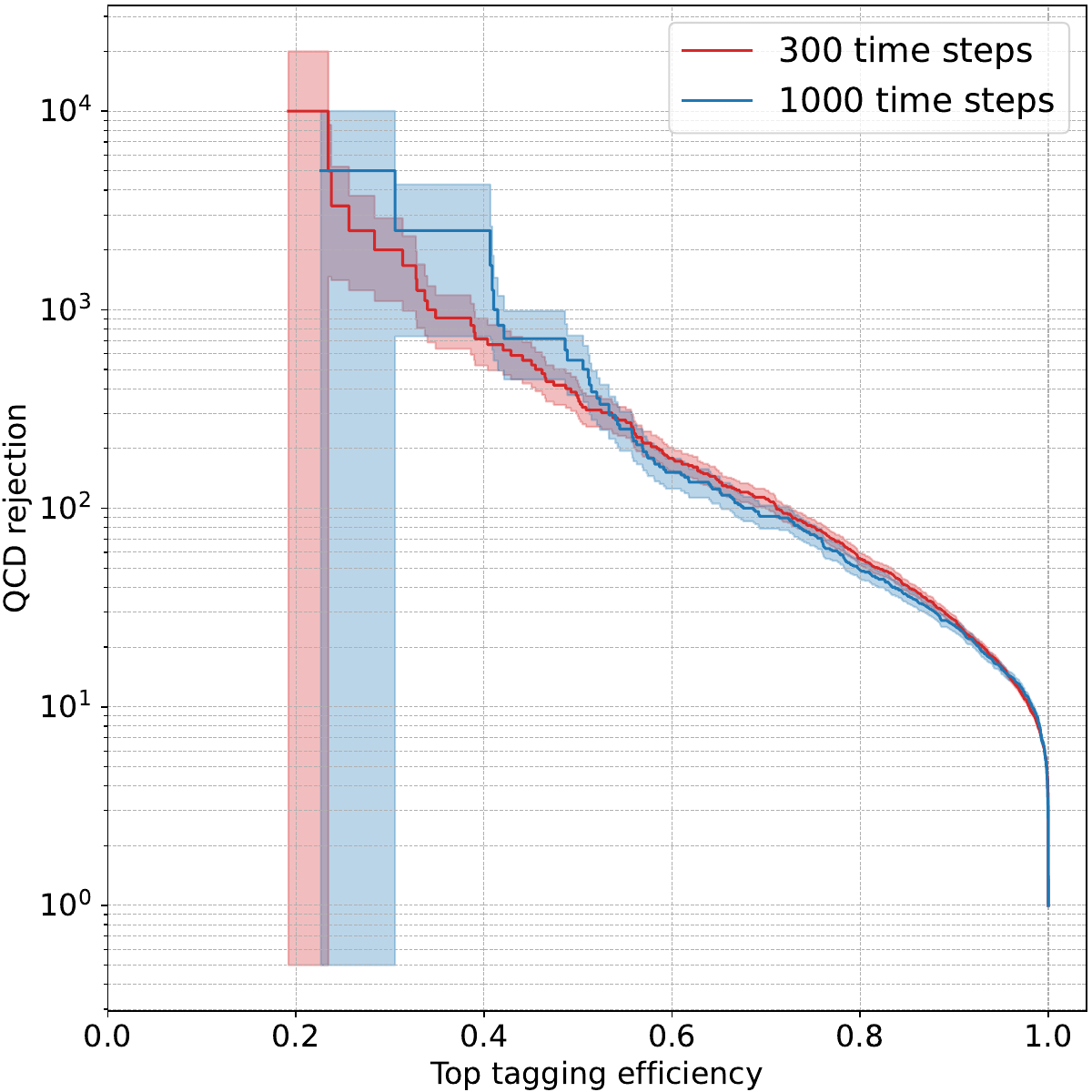}
    \caption{Comparison of NP-optimal ROC curves with number of inference/sampling timesteps. The blue curve shows the performance when both inference and sample generation use 300 time steps, while the red curve corresponds to 1000 time steps. The shaded bands denote binomial counting errors on the background sample and propagated to the QCD rejection axis.}
    \label{fig:timestep_compare}
\end{figure}

\subsubsection{Number of time steps}
To assess the dependence of the approximate Neyman-Pearson classifier performance on the number of time steps used for sample generation and inference of the log-likelihood values, we compared two configurations -- 300 versus 1000 time steps. For each configuration, the same number of time steps was used for sampling and inference. No fixed-point iteration refinement was used in either configuration. We found that using 1000 time steps for generation results in slightly higher fidelity jets. Nevertheless, as shown in Figure~\ref{fig:timestep_compare}, both configurations yield statistically consistent classification performance. The larger error bands for the 1000-step case are from the smaller evaluation sample size -- imposed by computational constraints -- used to construct the ROC curve. For our main results in the paper, we used the 300-step configuration in order to obtain a large training dataset for the jet taggers while keeping computational costs manageable.

Beyond these internal consistency checks, we also verify correctness in a controlled toy setting where the analytic NP ROC is known. This provides an end-to-end demonstration that FM likelihood ratios reproduce exact NP behavior in a case with closed-form ground truth.

\begin{figure}
\includegraphics[width=0.9\columnwidth]{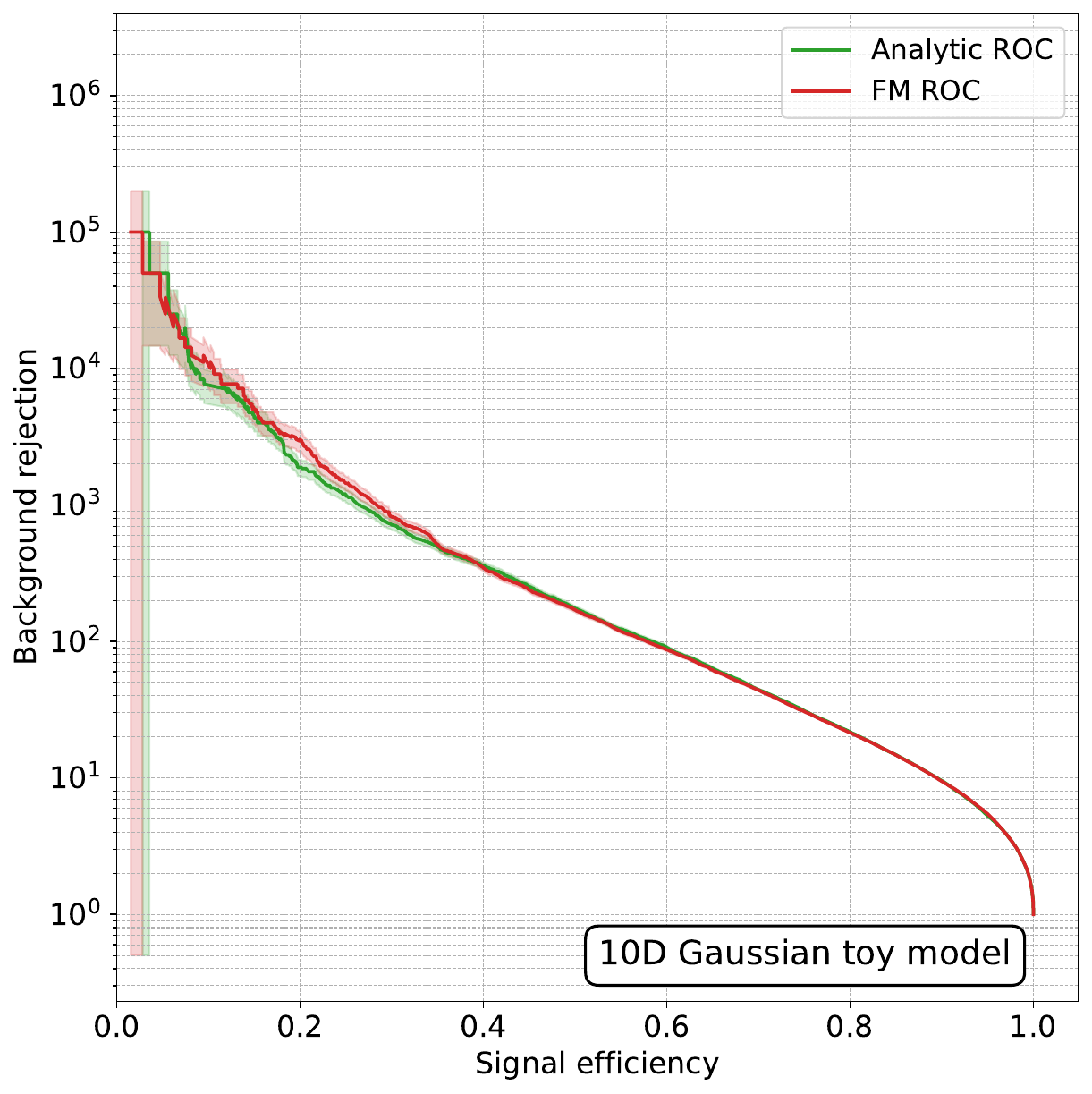}
    \caption{ROC curves for 10D gaussian toy example used for flow matching log-likelihood validation. The shaded bands denote binomial counting errors on the background sample and propagated to the background rejection axis.}
    \label{fig:gaussian_toy_roc}
\end{figure}

\subsubsection{Validation on a toy 10D Gaussian model}
\label{sec:toy} 
We construct two ten-dimensional Gaussian distributions with shared unit covariance but shifted means:
\[
p_{\text{sig}} = \mathcal{N}\!\left(+0.4\,\mathbf{1}_{10},\, I_{10}\right),
\quad
p_{\text{bg}}  = \mathcal{N}\!\left(-0.4\,\mathbf{1}_{10},\, I_{10}\right).
\]

In this setting, the NP test has a closed-form solution, and the corresponding ROC curve can be derived analytically. 
We train separate FM models, implemented as simple MLP-based vector-field networks, on samples from each Gaussian. After training, we generate samples from the FM models and evaluate the NP log-likelihood ratio using the FM log-density estimators on these generated samples.
Figure~\ref{fig:gaussian_toy_roc} shows that the ROC curve obtained from FM likelihood ratios coincides quite well with the analytic NP ROC and is mostly within statistical fluctuations. 

This toy test confirms that flow matching recovers the correct ROC behavior in a regime where the ground truth NP statistic is exactly known, supporting the validity of our approach before applying it to more complex jet data.

Taken together, these checks validate that FM-based log-likelihood ratios are numerically stable under integrator choices and, in the Gaussian toy model, reproduce the exact NP ROC. They do not claim absolute calibration of log-densities; rather, they establish that the relative log-densities that define our ROC curves are robust for the analyses presented here.

\section{Details of the GPT model}
\label{app:models-gpt}

For our model, we use the \texttt{GPT2LMHeadModel} implementation of the GPT-2 transformer architecture~\cite{Radford2019LanguageMA} provided by HuggingFace~\cite{wolf2020transformers}, configured with an initial token embedding to $\mathbb{R}^{d_{\rm embd}}$, followed by $n_{\rm layer}=12$ transformer blocks, each with $n_{\rm head}=8$ attention heads. We set $d_{\rm embd}=256$ for the embedding/hidden size, $d_{\rm inner}=1024$ for the feed-forward hidden size, $p_{\rm drop}=0.1$ for all dropout layers and {\tt GELU} for all non-linear activation functions. This resulting model contains approximately $16$ million trainable parameters. In addition to the standard causal mask that enforces autoregressive ordering, we apply an attention mask to ignore \texttt{[PAD]} tokens. The final ``language head" then maps the last hidden states to vocabulary logits through a linear layer. 

The model is trained by maximizing the autoregressive likelihood
(\ref{eq:gpt_ll}) with $\t t_0\equiv\texttt{[START]}$, excluding \texttt{[PAD]} tokens. Unlike EPiC\text{-}FM, which conditions on jet type, we train separate class-specific GPT models -- one for top jets and one for QCD jets. Optimization uses {\tt Adam}~\cite{kingma2014adam} with an effective batch size of 256 jets for up to 150 epochs: a cosine-annealing schedule decays the learning rate from $10^{-3}$ to $10^{-4}$ over the first 100 epochs, followed by 50 epochs at a fixed rate of $10^{-4}$. The final model is selected by the lowest validation loss.

\section{Results for $N_{\rm max} =128$}
\label{app:128particles}

In this appendix, we present a minimal extension of our study to higher particle multiplicity, increasing the maximum number of constituents per jet from $N_{\max}=40$ to $N_{\max}=128$. For this study, the EPiC-FM model is trained on \textsc{JetClass} jets in their continuous representation including up to the 128 hardest constituents, while the GPT model is trained on the corresponding tokenized representation. In order to maintain generation quality at this higher dimensionality, we increase the capacity of the EPiC-FM model accordingly.

\begin{figure*}
    \includegraphics[width=1.5\columnwidth]{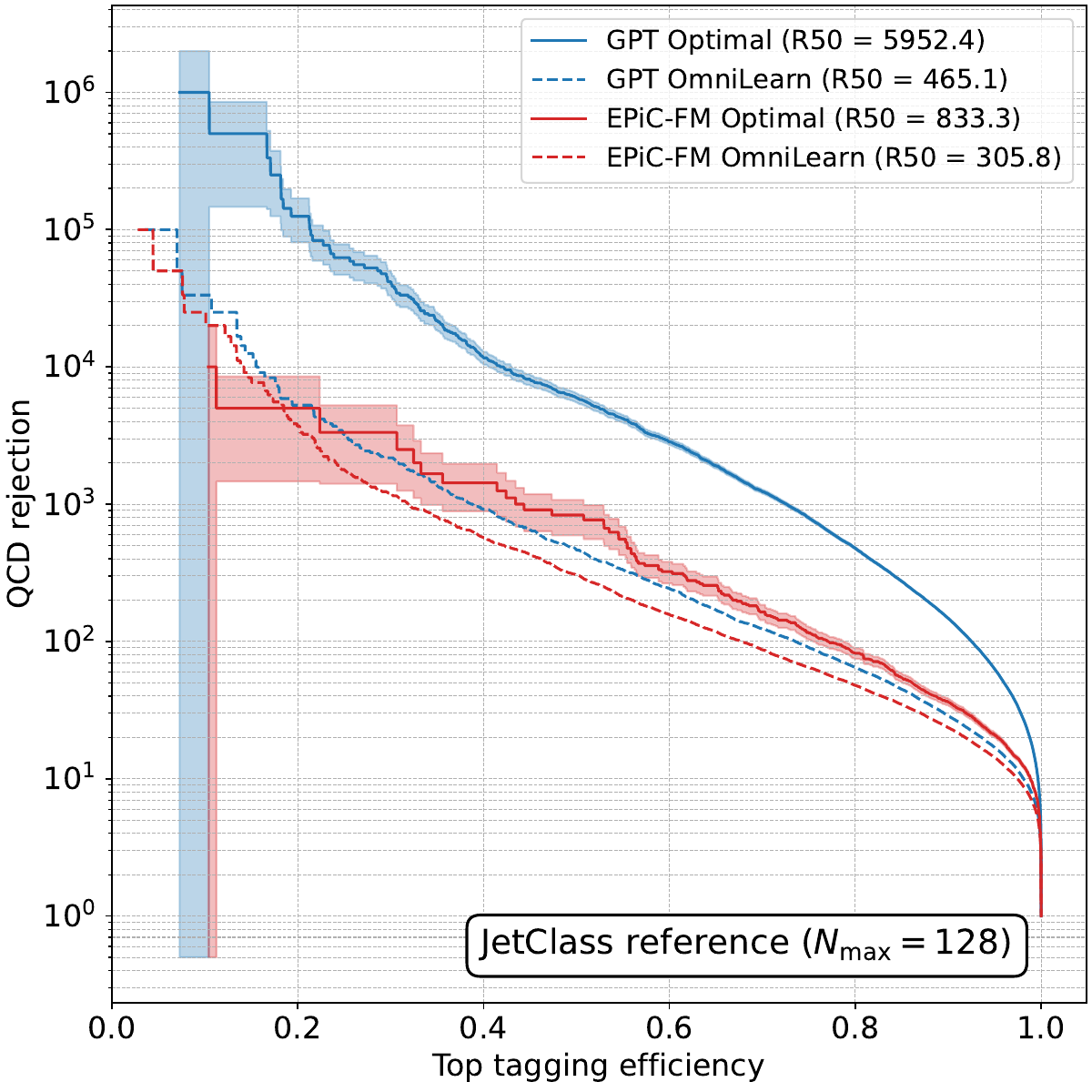}
    \caption{ROC curves for top vs.\ QCD jet classification derived from the \textbf{\textsc{JetClass}} dataset and the generative models trained on it. In contrast to the main study, each jet is represented by its 128 hardest constituents (i.e.~$N_\text{max} = 128$). Solid lines indicate the performance of the NP-optimal classifier obtained directly from the true log-likelihood ratio. Dashed lines correspond to classifiers trained with OmniLearn. The QCD rejection at a top tagging efficiency of 50\% (R50) is shown in parentheses for each curve. Shaded bands indicate the statistical uncertainty on the optimal ROC curves, estimated from binomial counting errors on the background sample and propagated to the QCD rejection axis.}
    \label{fig:jetclassref_optimalroc_nc128}
\end{figure*}

Figure~\ref{fig:jetclassref_optimalroc_nc128} shows the corresponding ROC curves for top vs.\ QCD classification. As in the main text, we compare the performance of the NP-optimal classifier, obtained from the log-likelihood ratio, with that of classifiers trained using OmniLearn.

We observe behavior consistent with our findings at $N_{\max}=40$. In particular, the GPT-based models exhibit a sizable discrepancy between the NP-optimal and classifier ROC curves, while the EPiC-FM NP-optimal ROC curve remains in closer agreement with the classifier performance.

These results support the robustness of our analysis with respect to particle multiplicity and indicate that restricting to $N_{\max}=40$ captures the essential features relevant for our study. Given the consistency of our findings, we do not find it necessary to perform a more extensive exploration of higher multiplicities.

\bibliography{HEPML,other-refs}

@preprint{Bhimji:2025isp,
    author = "Bhimji, Wahid and Harris, Chris and Mikuni, Vinicius and Nachman, Benjamin",
    title = "{OmniLearned: A Foundation Model Framework for All Tasks Involving Jet Physics}",
    eprint = "2510.24066",
    archivePrefix = "arXiv",
    primaryClass = "hep-ph",
    month = "10",
    year = "2025"
}

@article{Mikuni:2025tar,
    author = "Mikuni, Vinicius and Nachman, Benjamin",
    title = "{A Method to Simultaneously Facilitate All Jet Physics Tasks}",
    eprint = "2502.14652",
    journal="Phys.Rev.D",
    volume="111",
    pages="054015",
    doi="10.1103/PhysRevD.111.054015",
    archivePrefix = "arXiv",
    primaryClass = "hep-ph",
    month = "2",
    year = "2025"
}

@inproceedings{Geuskens:2024tfo,
    author = {Geuskens, Joep and Gite, Nishank and Kr\"amer, Michael and Mikuni, Vinicius and M\"uck, Alexander and Nachman, Benjamin and Reyes-Gonz\'alez, Humberto},
    title = "{The Fundamental Limit of Jet Tagging}",
    eprint = "2411.02628",
    archivePrefix = "arXiv",
    primaryClass = "hep-ph",
    month = "11",
    year = "2024"
}

@article{Brehmer:2024yqw,
    author = "Brehmer, Johann and Bres{\'o}, V{\'\i}ctor and de Haan, Pim and Plehn, Tilman and Qu, Huilin and Spinner, Jonas and Thaler, Jesse",
    title = "{A Lorentz-equivariant transformer for all of the LHC}",
    eprint = "2411.00446",
    archivePrefix = "arXiv",
    primaryClass = "hep-ph",
    reportNumber = "MIT-CTP/5802",
    doi = "10.21468/SciPostPhys.19.4.108",
    journal = "SciPost Phys.",
    volume = "19",
    number = "4",
    pages = "108",
    year = "2025"
}

@article{Cappelli:2025myc,
    author = "Cappelli, Pietro and Grosso, Gaia and Letizia, Marco and Reyes-Gonz{\'a}lez, Humberto and Zanetti, Marco",
    title = "{Learning to validate generative models: a goodness-of-fit approach}",
    eprint = "2511.09118",
    archivePrefix = "arXiv",
    primaryClass = "stat.ML",
    reportNumber = "TTK-25-36",
    doi = "10.1088/2632-2153/ae73e3",
    journal = "Mach. Learn. Sci. Tech.",
    volume = "7",
    number = "4",
    pages = "045011",
    year = "2026"
}

@preprint{Larkoski:2024hfe,
    author = "Larkoski, Andrew J.",
    title = "{Systematic Interpretability and the Likelihood for Boosted Top Quark Identification}",
    eprint = "2411.00104",
    archivePrefix = "arXiv",
    primaryClass = "hep-ph",
    month = "10",
    year = "2024"
}

@article{Krause:2024avx,
    author = "Amram, Oz and others",
    editor = "Krause, Claudius and Faucci Giannelli, Michele and Kasieczka, Gregor and Nachman, Benjamin and Salamani, Dalila and Shih, David and Zaborowska, Anna",
    title = "{CaloChallenge 2022: A Community Challenge for Fast Calorimeter Simulation}",
    eprint = "2410.21611",
    archivePrefix = "arXiv",
    primaryClass = "physics.ins-det",
    journal={Reports on Progress in Physics},
    reportNumber = "HEPHY-ML-24-05, FERMILAB-PUB-24-0728-CMS, TTK-24-43",
    doi = "10.1088/1361-6633/ae1304",
    month = "10",
    year = "2024"
}

@article{ATLAS:2024rua,
    author = "{ATLAS Collaboration}",
    title = "{Accuracy versus precision in boosted top tagging with the ATLAS detector}",
    eprint = "2407.20127",
    journal="JINST",
    volume="19",
    pages="P08018",
    doi="10.1088/1748-0221/19/08/P08018",
    archivePrefix = "arXiv",
    primaryClass = "hep-ex",
    reportNumber = "CERN-EP-2024-159",
    month = "7",
    year = "2024"
}

@article{Mikuni:2024qsr,
    author = "Mikuni, Vinicius and Nachman, Benjamin",
    title = "{Solving key challenges in collider physics with foundation models}",
    eprint = "2404.16091",
    archivePrefix = "arXiv",
    primaryClass = "hep-ph",
    doi = "10.1103/PhysRevD.111.L051504",
    journal = "Phys. Rev. D",
    volume = "111",
    number = "5",
    pages = "L051504",
    year = "2025"
}

@preprint{Favaro:2025pgz,
    author = "Favaro, Luigi and Gerhartz, Gerrit and Hamprecht, Fred A. and Lippmann, Peter and Pitz, Sebastian and Plehn, Tilman and Qu, Huilin and Spinner, Jonas",
    title = "{Lorentz-Equivariance without Limitations}",
    eprint = "2508.14898",
    archivePrefix = "arXiv",
    primaryClass = "hep-ph",
    month = "8",
    year = "2025"
}

@article{Birk:2024knn,
    author = "Birk, Joschka and Hallin, Anna and Kasieczka, Gregor",
    title = "{OmniJet-$\alpha$: The first cross-task foundation model for particle physics}",
    eprint = "2403.05618",
    journal="Mach.Learn.Sci.Tech.",
    volume="5",
    pages="035031",
    doi="10.1088/2632-2153/ad66ad",
    archivePrefix = "arXiv",
    primaryClass = "hep-ph",
    month = "3",
    year = "2024"
}

@preprint{Buhmann:2023zgc,
    author = "Buhmann, Erik and Ewen, Cedric and Faroughy, Darius A. and Golling, Tobias and Kasieczka, Gregor and Leigh, Matthew and Qu\'etant, Guillaume and Raine, John Andrew and Sengupta, Debajyoti and Shih, David",
    title = "{EPiC-ly Fast Particle Cloud Generation with Flow-Matching and Diffusion}",
    eprint = "2310.00049",
    archivePrefix = "arXiv",
    primaryClass = "hep-ph",
    month = "9",
    year = "2023"
}

@article{Golling:2024abg,
    author = "Golling, Tobias and Heinrich, Lukas and Kagan, Michael and Klein, Samuel and Leigh, Matthew and Osadchy, Margarita and Raine, John Andrew",
    title = "{Masked particle modeling on sets: towards self-supervised high energy physics foundation models}",
    eprint = "2401.13537",
    archivePrefix = "arXiv",
    primaryClass = "hep-ph",
    doi = "10.1088/2632-2153/ad64a8",
    journal = "Mach. Learn. Sci. Tech.",
    volume = "5",
    number = "3",
    pages = "035074",
    year = "2024"
}

@preprint{Bogatskiy:2023nnw,
    author = "Bogatskiy, Alexander and Hoffman, Timothy and Miller, David W. and Offermann, Jan T. and Liu, Xiaoyang",
    title = "{Explainable Equivariant Neural Networks for Particle Physics: PELICAN}",
    eprint = "2307.16506",
    journal="JHEP",
    volume="03",
    pages="113",
    doi="10.1007/JHEP03(2024)113",
    archivePrefix = "arXiv",
    primaryClass = "hep-ph",
    month = "7",
    year = "2023"
}

@article{Das:2023ktd,
    author = "Das, Ranit and Favaro, Luigi and Heimel, Theo and Krause, Claudius and Plehn, Tilman and Shih, David",
    title = "{How to Understand Limitations of Generative Networks}",
    eprint = "2305.16774",
    journal="SciPost Phys.",
    volume="16",
    pages="031",
    doi="10.21468/SciPostPhys.16.1.031",
    archivePrefix = "arXiv",
    primaryClass = "hep-ph",
    month = "5",
    year = "2023"
}

@article{Finke:2023veq,
    author = {Finke, Thorben and Kr\"amer, Michael and M\"uck, Alexander and T\"onshoff, Jan},
    title = "{Learning the language of QCD jets with transformers}",
    eprint = "2303.07364",
      journal="JHEP",
      volume="06",
      pages="184",
      doi="10.1007/JHEP06(2023)184",
    archivePrefix = "arXiv",
    primaryClass = "hep-ph",
    month = "3",
    year = "2023"
}

@article{Birk:2023efj,
    author = "Birk, Joschka and Buhmann, Erik and Ewen, Cedric and Kasieczka, Gregor and Shih, David",
    title = "{Flow matching beyond kinematics: Generating jets with particle identification and trajectory displacement information}",
    eprint = "2312.00123",
    archivePrefix = "arXiv",
    primaryClass = "hep-ph",
    doi = "10.1103/PhysRevD.111.052008",
    journal = "Phys. Rev. D",
    volume = "111",
    number = "5",
    pages = "052008",
    year = "2025"
}

@article{Buhmann:2023pmh,
    author = "Buhmann, Erik and Kasieczka, Gregor and Thaler, Jesse",
    title = "{EPiC-GAN: Equivariant Point Cloud Generation for Particle Jets}",
    eprint = "2301.08128",
      journal="SciPost Phys.",
      volume="15",
      pages="130",
      doi="10.21468/SciPostPhys.15.4.130",
    archivePrefix = "arXiv",
    primaryClass = "hep-ph",
    reportNumber = "MIT-CTP 5519",
    month = "1",
    year = "2023"
}

@article{Kansal:2022spb,
    author = "Kansal, Raghav and Li, Anni and Duarte, Javier and Chernyavskaya, Nadezda and Pierini, Maurizio and Orzari, Breno and Tomei, Thiago",
    title = "{Evaluating generative models in high energy physics}",
    eprint = "2211.10295",
    archivePrefix = "arXiv",
    primaryClass = "hep-ex",
    reportNumber = "FERMILAB-PUB-22-872-CMS-PPD",
    doi = "10.1103/PhysRevD.107.076017",
    journal = "Phys. Rev. D",
    volume = "107",
    number = "7",
    pages = "076017",
    year = "2023"
}

@preprint{Qu:2022mxj,
    author = "Qu, Huilin and Li, Congqiao and Qian, Sitian",
    title = "{Particle Transformer for Jet Tagging}",
    eprint = "2202.03772",
    archivePrefix = "arXiv",
    primaryClass = "hep-ph",
    month = "2",
    year = "2022"
}

@article{Gong:2022lye,
    author = "Gong, Shiqi and Meng, Qi and Zhang, Jue and Qu, Huilin and Li, Congqiao and Qian, Sitian and Du, Weitao and Ma, Zhi-Ming and Liu, Tie-Yan",
    title = "{An Efficient Lorentz Equivariant Graph Neural Network for Jet Tagging}",
    eprint = "2201.08187",
      journal="JHEP",
      volume="07",
      pages="030",
      doi="10.1007/JHEP07(2022)030",
    archivePrefix = "arXiv",
    primaryClass = "hep-ph",
    month = "1",
    year = "2022"
}

@article{Krause:2021ilc,
    author = "Krause, Claudius and Shih, David",
    title = "{CaloFlow: Fast and Accurate Generation of Calorimeter Showers with Normalizing Flows}",
    eprint = "2106.05285",
      journal="Phys.Rev.D",
      volume="107",
      pages="113003",
      doi="10.1103/PhysRevD.107.113003",
    archivePrefix = "arXiv",
    primaryClass = "physics.ins-det",
    month = "6",
    year = "2021"
}

@article{Komiske:2018cqr,
    author = "Komiske, Patrick T. and Metodiev, Eric M. and Thaler, Jesse",
    archivePrefix = "arXiv",
    doi = "10.1007/JHEP01(2019)121",
    eprint = "1810.05165",
    journal = "JHEP",
    pages = "121",
    primaryClass = "hep-ph",
    reportNumber = "MIT-CTP 5064",
    title = "{Energy Flow Networks: Deep Sets for Particle Jets}",
    volume = "01",
    year = "2019"
}

@article{Qu:2019gqs,
    author = "Qu, Huilin and Gouskos, Loukas",
    archivePrefix = "arXiv",
    doi = "10.1103/PhysRevD.101.056019",
    eprint = "1902.08570",
    journal = "Phys. Rev. D",
    number = "5",
    pages = "056019",
    primaryClass = "hep-ph",
    title = "{ParticleNet: Jet Tagging via Particle Clouds}",
    volume = "101",
    year = "2020"
}

@article{Kasieczka:2019dbj,
    author = "Butter, Anja and others",
    editor = "Kasieczka, Gregor and Plehn, Tilman",
    archivePrefix = "arXiv",
    doi = "10.21468/SciPostPhys.7.1.014",
    eprint = "1902.09914",
    journal = "SciPost Phys.",
    pages = "014",
    primaryClass = "hep-ph",
    title = "{The Machine Learning Landscape of Top Taggers}",
    volume = "7",
    year = "2019"
}

@article{lipman2022flow,
  title={Flow matching for generative modeling},
  author={Lipman, Yaron and Chen, Ricky TQ and Ben-Hamu, Heli and Nickel, Maximilian and Le, Matt},
  journal={arXiv preprint arXiv:2210.02747},
  year={2022}
}

@article{albergo2022building,
  title={Building normalizing flows with stochastic interpolants},
  author={Albergo, Michael S and Vanden-Eijnden, Eric},
  journal={arXiv preprint arXiv:2209.15571},
  year={2022}
}

@article{CMS:2020poo,
    author = "Sirunyan, Albert M and others",
    collaboration = "CMS",
    title = "{Identification of heavy, energetic, hadronically decaying particles using machine-learning techniques}",
    eprint = "2004.08262",
    archivePrefix = "arXiv",
    primaryClass = "hep-ex",
    reportNumber = "CMS-JME-18-002, CERN-EP-2020-037",
    doi = "10.1088/1748-0221/15/06/P06005",
    journal = "JINST",
    volume = "15",
    number = "06",
    pages = "P06005",
    year = "2020"
}

@article{tong2023improving,
  title={Improving and generalizing flow-based generative models with minibatch optimal transport},
  author={Tong, Alexander and Fatras, Kilian and Malkin, Nikolay and Huguet, Guillaume and Zhang, Yanlei and Rector-Brooks, Jarrid and Wolf, Guy and Bengio, Yoshua},
  journal={arXiv preprint arXiv:2302.00482},
  year={2023}
}

@dataset{JetClass,
  author       = "Qu, Huilin and Li, Congqiao and Qian, Sitian",
  title        = "{JetClass}: A Large-Scale Dataset for Deep Learning in Jet Physics",
  month        = "jun",
  year         = "2022",
  publisher    = "Zenodo",
  version      = "1.0.0",
  doi          = "10.5281/zenodo.6619768",
  url          = "https://doi.org/10.5281/zenodo.6619768"
}

@inproceedings{Radford2019LanguageMA,
  title={Language Models are Unsupervised Multitask Learners},
  author={Alec Radford and Jeff Wu and Rewon Child and David Luan and Dario Amodei and Ilya Sutskever},
  year={2019},
  url={https://api.semanticscholar.org/CorpusID:160025533}
}

@article{Neyman:1933wgr,
    author = "Neyman, Jerzy and Pearson, Egon Sharpe",
    title = "{On the Problem of the Most Efficient Tests of Statistical Hypotheses}",
    doi = "10.1098/rsta.1933.0009",
    journal = "Phil. Trans. Roy. Soc. Lond. A",
    volume = "231",
    number = "694-706",
    pages = "289--337",
    year = "1933"
}

@article{Lim:2022nft,
    author = "Lim, Sung Hak and Raman, Kailash A. and Buckley, Matthew R. and Shih, David",
    title = "{GalaxyFlow: upsampling hydrodynamical simulations for realistic mock stellar catalogues}",
    eprint = "2211.11765",
    archivePrefix = "arXiv",
    primaryClass = "astro-ph.GA",
    doi = "10.1093/mnras/stae1672",
    journal = "Mon. Not. Roy. Astron. Soc.",
    volume = "533",
    number = "1",
    pages = "143--164",
    year = "2024"
}

@article{zhang2023diffusion,
  title={Diffusion noise feature: Accurate and fast generated image detection},
  author={Zhang, Yichi and Xu, Xiaogang},
  journal={arXiv preprint arXiv:2312.02625},
  year={2023}
}

@article{zaheer2017deep,
  title={Deep sets},
  author={Zaheer, Manzil and Kottur, Satwik and Ravanbakhsh, Siamak and Poczos, Barnabas and Salakhutdinov, Russ R and Smola, Alexander J},
  journal={Advances in neural information processing systems},
  volume={30},
  year={2017}
}

@inproceedings{rahaman2019spectral,
  title={On the spectral bias of neural networks},
  author={Rahaman, Nasim and Baratin, Aristide and Arpit, Devansh and Draxler, Felix and Lin, Min and Hamprecht, Fred and Bengio, Yoshua and Courville, Aaron},
  booktitle={International conference on machine learning},
  pages={5301--5310},
  year={2019},
  organization={PMLR}
}

@dataset{reyes_gonzalez_2024_14023638,
  author       = {Reyes-Gonzalez, Humberto and
                  Krämer, Michael and
                  Mück, Alexander and
                  Nachman, Benjamin and
                  Mikuni, Vinicius and
                  Geuskens, Joep and
                  Gite, Nishank},
  title        = {Resources for The Fundamental Limit of Jet Tagging},
  month        = nov,
  year         = 2024,
  publisher    = {Zenodo},
  doi          = {10.5281/zenodo.14023638},
  url          = {https://doi.org/10.5281/zenodo.14023638},
}

@article{kadkhodaie2023generalization,
  title={Generalization in diffusion models arises from geometry-adaptive harmonic representations},
  author={Kadkhodaie, Zahra and Guth, Florentin and Simoncelli, Eero P and Mallat, St{\'e}phane},
  journal={arXiv preprint arXiv:2310.02557},
  year={2023}
}

@article{vastola2025generalization,
  title={Generalization through variance: how noise shapes inductive biases in diffusion models},
  author={Vastola, John J},
  journal={arXiv preprint arXiv:2504.12532},
  year={2025}
}

@article{bertrand2025closed,
  title={On the Closed-Form of Flow Matching: Generalization Does Not Arise from Target Stochasticity},
  author={Bertrand, Quentin and Gagneux, Anne and Massias, Mathurin and Emonet, R{\'e}mi},
  journal={arXiv preprint arXiv:2506.03719},
  year={2025}
}

@article{ronen2019convergence,
  title={The convergence rate of neural networks for learned functions of different frequencies},
  author={Ronen, Basri and Jacobs, David and Kasten, Yoni and Kritchman, Shira},
  journal={Advances in Neural Information Processing Systems},
  volume={32},
  year={2019}
}

@article{press2017using,
  title={Using the Output Embedding to Improve Language Models},
  author={Press, Ofir and Wolf, Lior},
  journal={Proceedings of the 15th Conference of the European Chapter of the Association for Computational Linguistics},
  year={2017}
}

@inproceedings{inan2017tying,
  title={Tying Word Vectors and Word Classifiers: A Loss Framework for Language Modeling},
  author={Inan, Hakan and Khosravi, Khashayar and Socher, Richard},
  booktitle={International Conference on Learning Representations (ICLR)},
  year={2017}
}

@inproceedings{wolf2020transformers,
  title={Transformers: State-of-the-Art Natural Language Processing},
  author={Wolf, Thomas and Debut, Lysandre and Sanh, Victor and Chaumond, Julien and Delangue, Clement and Moi, Anthony and Cistac, Pierric and Rault, Tim and Louf, Remi and Funtowicz, Morgan and Davison, Joe and Shleifer, Sam and von Platen, Patrick and Ma, Clara and Jernite, Yacine and Plu, Julien and Xu, Canwen and Scao, Teven Le and Gugger, Sylvain and Drame, Mariama and Lhoest, Quentin and Rush, Alexander M},
  booktitle={Proceedings of the 2020 Conference on Empirical Methods in Natural Language Processing: System Demonstrations},
  pages={38--45},
  year={2020},
  organization={Association for Computational Linguistics}
}

@article{kingma2014adam,
  title={Adam: A method for stochastic optimization},
  author={Kingma, Diederik P and Ba, Jimmy},
  journal={arXiv preprint arXiv:1412.6980},
  year={2014}
}

\end{document}